\documentclass[3p,preview]{elsarticle}
\biboptions{sort&compress}
\usepackage{booktabs}

\usepackage{amsmath} 
\usepackage{multirow}
\usepackage{xr}
\makeatletter
\newcommand*{\addFileDependency}[1]{
  \typeout{(#1)}
  \@addtofilelist{#1}
  \IfFileExists{#1}{}{\typeout{No file #1.}}
}
\makeatother

\newcommand*{\myexternaldocument}[1]{%
    \externaldocument{#1}%
    \addFileDependency{#1.tex}%
    \addFileDependency{#1.aux}%
}
\myexternaldocument{SupplementaryMaterial}

\begin{document}
\begin{frontmatter}

\title{Resolving the Metastable Si-XIII Structure \\through Convergent Theory and Experiment}

\author[1]{Fabrizio Rovaris\fnref{fn1}}
\author[4]{Corrado Bongiorno\fnref{fn1}}
\author[1]{Anna Marzegalli}
\author[1,7]{Mouad Bikerouin}
\author[6]{Davide Spirito}
\author[3]{Gerald J. K. Schaffar}
\author[4]{Mohamed Zaghloul}
\author[8]{Agnieszka Anna Corley-Wiciak}
\author[1]{Francesco Montalenti}
\author[3]{Verena Maier-Kiener}
\author[2,5]{Giovanni Capellini\corref{cor1}}
\ead{capellini@ihp-microelectronics.com}
\author[4]{Antonio M. Mio}
\author[1]{Emilio Scalise\corref{cor1}}
\ead{emilio.scalise@unimib.it}

\affiliation[1]{organization={Department of Materials Science. University of Milano-Bicocca},
                addressline={Via R. Cozzi 55},
                postcode={20125},
                city={Milano},
                country={Italy}}

\affiliation[2]{organization={IHP-Leibniz Institute for High Performance Microelectronics},   
                addressline={Im Technologiepark 25}, 
                postcode={15236}, 
                city={Frankfurt(Oder)}, 
                country={Germany}}

\affiliation[3]{organization={Department of Materials Science. Montanuniversität Leoben},                                    addressline={Roseggerstrasse 12}, 
                postcode={8700}, 
                city={Leoben}, 
                country={Austria}}

\affiliation[4]{organization={Institute for Microelectronics and Microsystems (IMM). Consiglio Nazionale delle Ricerche (CNR)}, 
                addressline={Strada VIII 5}, 
                postcode={95121}, 
                city={Catania}, 
                country={Italy}}
                
\affiliation[5]{organization={Department of Sciences. Università Roma Tre},
                addressline={V.le G. Marconi 446},
                postcode={00146},
                city={Rome},
                country={Italy}}

\affiliation[6]{organization={BCMaterials. Basque Center for Materials, Applications and Nanostructures},
                addressline={UPV/EHU Science Park},
                postcode={48940},
                city={Leioa},
                country={Spain}}
                
\affiliation[7]{organization={Aalto University, Engineered Nanosystems Group},
                addressline={P.O. Box 13500},
                postcode={FI-00076},
                city={Aalto}, 
                country={Finland}}

\affiliation[8]{organization={European Synchrotron Radiation Facility},
                addressline={Cedex 9},
                postcode={38043},
                city={Grenoble},
                country={France}}

\cortext[cor1]{Corresponding authors}
\fntext[fn1]{These authors equally contributed}

\begin{keyword}
Si-XIII \sep Metastable silicon allotropes \sep Nanoindentation \sep TEM/SAED \sep Raman spectroscopy \sep Phase Transition \sep Kinetics
\end{keyword}

\begin{abstract}
Silicon allotropes have emerged as a pivotal frontier for engineering materials with tailored optical, electronic, and thermal functionalities, opening routes toward integrated photonics, energy harvesting, and quantum technologies, while avoiding rare or toxic elements. High-pressure experiments have revealed several metastable silicon allotropes, among which Si-XIII stands out. First observed more than 20 years ago, this phase has remained structurally unidentified, representing a significant gap in our understanding of elemental silicon allotropy. In this work, a convergent methodology is employed combining advanced theoretical modeling with experimental characterization to finally resolve the long-standing structural assignment of Si-XIII. We construct a structural model guided by experimental observations, validate it through first-principles optimization, and systematically test it against multiple experimental signatures. All key fingerprints of this phase — interplanar spacings, Raman frequencies, thermodynamic stability, and kinetic pathways — are consistently rationalized by our proposed crystal structure. These findings provide a crucial missing piece in the high-pressure phase diagram of silicon, demonstrate the power of integrating computational predictions with experimental validation to resolve complex structural problems in materials science, and open new avenues for the controlled synthesis and functional exploitation of metastable Group-IV allotropes.
\end{abstract}

\end{frontmatter}

\section{Introduction}\label{sec:intro}

The exploration of allotropes and polymorphs has emerged as a powerful lever for engineering materials with tailored properties~\cite{Poly,Kim2023,Shao2024,ChenNatRev2020}. Group‑IV materials have been at the forefront of these efforts, with several new allotropes successfully synthesized and characterized in recent years~\cite{LaiNature2026,LuoNature2022,FadalyNature2020,ShiellPRL2021,Barth2020}, making them particularly attractive for materials engineering without relying on rare and potentially toxic elements. These allotropes are thermodynamically distinct from the ground state and accessible only under specific non-equilibrium conditions. In silicon, metastable phases are often obtained through high-pressure treatments like nanoindentation or diamond anvil cell compression~\cite{Bikerouin2025_SmallStruct_hdSi,Haberl2016_APR_ExoticSi}, which drive structural transformations away from the diamond cubic (\emph{dc}) ground state. Several non-diamond silicon allotropes have been predicted, and in some cases experimentally demonstrated, to exhibit reduced thermal conductivity~\cite{ShaoMatTodPhys2022}, enhanced thermoelectric performance~\cite{ZhangMSMSE2018,LiuEPJB2021}, high carrier mobility~\cite{ShenPCCP2022}, superconductivity~\cite{SungPRL2018}, or even a direct and quasi-direct fundamental bandgap~\cite{KimNatMat2015,WangJACS2014,HePCCP2016}. These properties open new technology routes for on-chip energy harvesting, integrated photonics, and more exotic quantum functionalities. However, pressure-induced transformations can also cause wafer failure, defect generation, and performance degradation. Understanding the stress-driven nucleation mechanisms of these allotropes is therefore crucial both for unlocking new processing possibilities and for enhancing material quality. Remarkably, pressure-induced phase transitions in silicon have unveiled a diverse array of metastable phases. Despite decades of experimental and theoretical research, some of these phases remain structurally uncharacterized. Among those, Si-XIII is the perfect epitome: a phase first reported over twenty years ago by Gogotsi and coworkers~\cite{domnich2002phase} on the basis of Raman and electron diffraction signatures inconsistent with any known silicon allotrope.

Identifying and decoding metastable phases is one of the most difficult problems in materials science. These phases usually appear in heterogeneous mixtures where the desired allotropes coexist with amorphous silicon, diamond-cubic remnants, or other metastable structures. These structures are often organized into misaligned nanometric grains~\cite{Bikerouin2025_SmallStruct_hdSi,MeghaAdvFunMat2025}.  This complexity makes interpreting experimental data like Raman spectra or diffraction patterns particularly difficult when the underlying atomic structure is unknown. Resolving these ambiguities requires a rigorous synergistic approach between theory and experiment, cross-validating multiple characterization techniques with advanced theoretical modeling. Such a convergent approach has been adopted here to finally resolve the crystalline structure of the long-debated Si-XIII phase. A bulk atomistic model is presented, which achieves comprehensive agreement with all available experimental evidence. This model is supported by a solid theoretical framework describing the kinetics of phase transitions to and from this structure, as well as its relationships with the stable diamond cubic (also known as Si-I) and hexagonal diamond (\emph{hd}, or Si-IV) phases, and the well-known high-pressure R8 (Si-XII) and BC8 (Si-III) allotropes~\cite{Bikerouin2025_SmallStruct_hdSi}.

Through nanoindentation-based mechanical processing and controlled thermal treatment, a phase exhibiting the characteristic Raman signature of Si-XIII~\cite{Haberl2016_APR_ExoticSi,Wong2019_JAP_SiXIII} has been obtained. To unambiguously identify this structure, meticulous transmission electron microscopy (TEM) and selected area electron diffraction (SAED) analysis were performed. These analyses cross-referenced diffraction data from multiple zone axes to rule out compatibility with other known silicon allotropes and to isolate the specific volume transformed into the new phase. Building on these rigorous diffraction-derived descriptions of the crystal lattice, a first-principles approach was employed to refine candidate structures, ultimately leading to the identification of a configuration that is crystallographically consistent, energetically metastable, and whose computed Raman spectrum accurately reproduces the unique experimental signature of Si-XIII.

Beyond static structural validation, an advanced implementation of the solid-state dimer method~\cite{Henkelman2000_JCP_dimer,PenghaoJCP2014}, specifically tailored for crystalline solids, was employed to map the kinetic pathways connecting Si-XIII to both diamond phases (diamond cubic and hexagonal diamond) as well as to the high-pressure R8 and BC8 metastable phases. The resulting transition pathways and energy barriers are fully consistent with the mechanical loading and annealing conditions used experimentally, embedding the proposed structure into a coherent thermodynamic and kinetic framework that provides a comprehensive atomistic picture of the stress- and temperature-driven phase transformation landscape of silicon.

\section{Methods}
\label{sec:methods}

\subsection{Nanoindentation}
Nanoindentation was performed on monocrystalline silicon (001) using a KLA Corporation G200 System (Milpitas, CA, USA) equipped with spherical diamond indenter tips of 10~$\mu$m and 20~$\mu$m radii (Synton-MDP, Nidau, Switzerland). Loading proceeded at a constant strain rate $\dot{\epsilon} = 10^{-3}$~s$^{-1}$~\cite{LEITNER2018, Kiener2023}, while unloading was conducted at a constant rate of $\dot{P} = 1$~mN/s to promote crystalline phase transformation over amorphization~\cite{lin2020temperature, domnich2002phase, Schaffar2022, Juliano2004}. For the 20~$\mu$m tip, the maximum load was set at $P_{\text{max}} = 665$~mN to ensure complete phase transformation, while with the 10~$\mu$m tip, indentations were performed at approximately $P_{\text{max}} = 425$~mN. Further experimental details are provided in Ref.~\cite{Bikerouin2025_SmallStruct_hdSi}.
Post-indentation thermal treatment was carried out on a temperature-controlled Raman stage under nitrogen flow. The annealing cycle comprised: (1) baseline measurement at 20~$^{\circ}$C, (2) heating to 220~$^{\circ}$C (50~$^{\circ}$C/min), (3) stepwise cooling to 20~$^{\circ}$C with Raman acquisition every 20~°C, (4) second heating to 220~°C (50~°C/min), and (5) final cooling to 20~°C (50~$^{\circ}$C/min). 
In a selected case, after completing the stage-based thermal cycling and Raman characterization, supplementary furnace annealing was performed at 250~°C for 2~h in a quartz tube under a nitrogen atmosphere, after which the sample was cooled to room temperature in air.

\subsection{TEM, SAED}
TEM and SAED were performed using a JEOL JEM-2010F microscope equipped with a Schottky field emission gun operating at 200~keV. Cross-sectional lamellae were prepared by focused ion beam (FIB) milling in a Thermo Scientific™ Helios™ 5 UC DualBeam system using 30~keV Ga$^+$ ions, followed by low-energy (2~keV Ga$^+$) polishing to minimize FIB-induced amorphization. Moreover, we prepared the TEM lamellae with an uneven thickness profile, with the cut on one side of the observed pit being thicker (about 300 nm) than the other side (about 100 nm). We chose this procedure to provide some mechanical support to the TEM lamellae, which reduced the extent of relaxation under thinning. Under these conditions, we did not observe any change under several TEM observations carried out over a period of several weeks, differently from some observations in the literature (see, e.g., Refs~\cite{Ge2004,Wong2019_JAP_SiXIII}, where the Si-XIII phase exhibited fast decaying metastability when reduced to thin lamellae.

In general, the uncertainty in the determination of \textit{d}-spacings obtained by SAED is typically on the order of 0.05 Å for spacings below 4 Å and can reach approximately 0.1 Å for larger spacings~\cite{Wong2019_JAP_SiXIII}. However, in systematic measurements, particularly when comparing reflections within the same diffraction pattern, the uncertainty is significantly reduced. In our experience, for \textit{d}-spacings around 3 Å, the uncertainty is better than 0.02 Å in a single crystal. This level of precision is consistently observed, for example, in measurements of the \{111\} reflections of diamond-cubic silicon. 

The electron diffraction patterns were interpreted using the JEMS version 4.13531u2024b31 software (JEMS-SAS, Pierre Stadelmann, Switzerland).

\subsection{Raman Spectroscopy}

Raman measurements were conducted using a Renishaw inVia system with 532~nm excitation in back-scattering configuration, with the incident laser aligned along the substrate's [001] crystallographic direction (z-axis). The setup features a polarizer for the excitation laser and an analyzer at the spectrometer's entrance. Additionally, an half-wave plate allows to rotate the polarization of the excitation by an angle $\Omega$. Samples were oriented with the substrate's $x' = [110]$ direction aligned to $\Omega = 0^{\circ}$. Polarization-resolved spectra were systematically acquired in both parallel ($\Omega = 0^{\circ}$) and perpendicular ($\Omega = 90^{\circ}$) configurations to track phase transformations and crystallographic evolution.

\subsection{Atomic Force Microscopy}

Atomic force microscopy (AFM) images were recorded using a Dimension ICON from Bruker in tapping mode and using Bruker TESPA VS-SS probes, featuring a nominal tip radius of approximately 5 nm.  Image analysis and rendering has been carried out using the freeware Gwyddion software available at Ref.~\cite{gwyddion}

\subsection{First-Principles Calculations}

Density Functional Theory (DFT) calculations were carried out with the Quantum ESPRESSO package~\cite{giannozzi2020quantum}. The exchange-correlation functional was treated primarily with the local density approximation (LDA)~\cite{kohn1965self}. For benchmarking purposes, the Perdew-Burke-Ernzerhof (PBE) generalized gradient approximation (GGA)~\cite{perdew1996generalized}, the strongly constrained and appropriately normed (SCAN) meta-GGA~\cite{sun2015strongly, yao2017plane} via the LIBXC library~\cite{lehtola2018recent}, and the PBEsol functional optimized for solids~\cite{perdew2008restoring} were also tested. 

Electron-ion interactions were described using norm-conserving Hartwigesen-Goedecker-Hutter~\cite{Goedecker1996, Hartwigsen1998} pseudopotentials for LDA, PBE, and SCAN calculations, while the PBEsol case employed a pseudopotential from the standard Quantum ESPRESSO library~\cite{giannozzi2020quantum}. A plane-wave basis set with a kinetic energy cutoff of 80~Ry was employed. Brillouin zone integration was performed using an 8$\times$8$\times$8 $\Gamma$-centered k-point mesh following the Monkhorst-Pack scheme~\cite{monkhorst1976special}.

Structural relaxations were performed using variable-cell optimization until residual forces fell below 1~meV/\AA\ and stress tensor components fell below 0.02~kbar, with self-consistent field convergence set to $10^{-10}$~eV. Dynamical stability and phonon dispersion relations were assessed via density functional perturbation theory (DFPT)~\cite{baroni2001phonons}. Non-resonant Raman spectra, including mode frequencies and scattering intensities, were computed within the DFPT framework as implemented in Quantum ESPRESSO.

The VESTA software~\cite{VESTA} and the Materials Project toolkit~\cite{MaterialsProj} were employed to display and manipulate crystal structures, while the crystal symmetry analysis was performed using the Materials Project toolkit~\cite{MaterialsProj} and the XRDlicious point defect submodule~\cite{XRDilicous}

\subsection{Potential Energy Surface Exploration}

Potential Energy Surface (PES) explorations were carried out by repeated iterations of the Solid State Dimer (SS-Dimer) method~\cite{PenghaoJCP2014}. We performed a total of about 100000 dimer searches with random initial displacements starting from the PES minima of each considered phase.  As it would be unfeasible to perform such an extended analysis within an ab-initio framework, PES calculations were performed exploiting the GAP machine-learning interatomic potential for Si~\cite{BartokPRX2018}, which was recently shown to be  particularly reliable in describing the kinetics of phase transitions in silicon~\cite{RovarisMTN2025,GeActaMat2024}. Initially, we optimised all relevant Si phases including the candidate Si-XIII structure.  The thermodynamic properties predicted by the GAP potential and DFT, particularly using the PBE functional, showed excellent agreement as demonstrated in Table S2 in the Supplementary Material file.

More technically, we initialized the dimer iterations by performing small displacements of the atomic positions only, the cell parameters only, and both combined. The amplitude of the displacements was varied, sampling from a normal distribution with a standard deviation in the range of $[0.01-0.3]~\text{\AA}$ to sample in the most uniform and complete way the neighborhood of the PES. We selected a maximum energy threshold for saddle point location of $300~\text{meV/atom}$. To avoid artifacts arising from the dimer being displaced from the initial energy basin, we energy-optimized the two replicas in the converged dimer for each saddle point found to ensure connectivity with the initial energy minimum. The resulting kinetic database was then analyzed and ordered by energy barrier values to highlight the most physically-relevant transitions. Specific paths were also re-evaluated by Solid State Nudged Elastic Band (SS-NEB)~\cite{SheppardJCP2012} calculations initialized with the saddle point and the two minima found by the SS-Dimer calculation. This allowed us to draw the Minimum Energy Paths and evaluate the transition images in between the two connected minima.
The energy convergence for saddle point location and for the minimization of the Dimer images was set to $10^{-3}~\text{eV\AA}$ and $10^{-4}~\text{eV/\AA}$, respectively. The TSASE implementation~\cite{tsase} for the Atomic Simulation Environment (ASE)~\cite{LarsenJPCM2017} of the SS-Dimer method was exploited in this work.

\section{Results and Discussion}
\label{sec:res}

\subsection{From SAED to the crystal structure of the Si-XIII phase}
\label{sec:res_SAED}

Numerous theoretical and experimental studies have attempted to identify and assign a crystal structure to the Si-XIII phase. However, it remains one of the few experimentally-identified silicon allotropes whose crystalline structure remains unknown~\cite{Wong2019_JAP_SiXIII}. Experimental evidence consistently associates Si-XIII with characteristic Raman peaks at approximately $475~\text{cm}^{-1}$ and $330~\text{cm}^{-1}$, together with electron-diffraction spots at $\sim 5.6~\text{\AA}$ and  $\sim 4.4~\text{\AA}$, which cannot be attributed to any other known silicon phase~\cite{domnich2002phase, Ruffell2009, Haberl2015, Haberl2016_APR_ExoticSi, Wong2019_JAP_SiXIII}. Despite extensive literature on this elusive phase, most studies have been limited to identifying individual plane spacings, leaving no definitive assignment to a crystal structure that fully reconciles all experimental observations.

Here, we address this challenge through comprehensive TEM and SAED investigations of nanoindented silicon subjected to ramped thermal annealing up to $220^{\circ}$C. Building on the observation by Wong et al.~\cite{Wong2019_JAP_SiXIII}, we found that short thermal annealing protocols at temperatures between $200^{\circ}$C and $240^{\circ}$C favor the transformation of the R8 metastable phase, formed upon indentation, into Si-XIII. Therefore, we employed a staged thermal treatment up to $220^{\circ}$C that produces distinct diffraction signatures consistent with the Si-XIII patterns previously reported in the literature~\cite{Haberl2015, Wong2019_JAP_SiXIII}. This approach contrasts with longer ($\geq$~2~h) isothermal treatments at temperatures above $\sim 240^{\circ}$C, which predominantly result in \emph{hd}~\cite{Bikerouin2025_SmallStruct_hdSi, Wong2019_JAP_SiXIII}. 
To extract reliable structural information from electron diffraction, analogous to single-crystal X-ray diffraction protocols, we isolated individual grains and systematically acquired multiple low-index zone axes from the same crystallite.

\begin{figure*}[h!tb]
    \centering
    \includegraphics[width=1.0\linewidth]{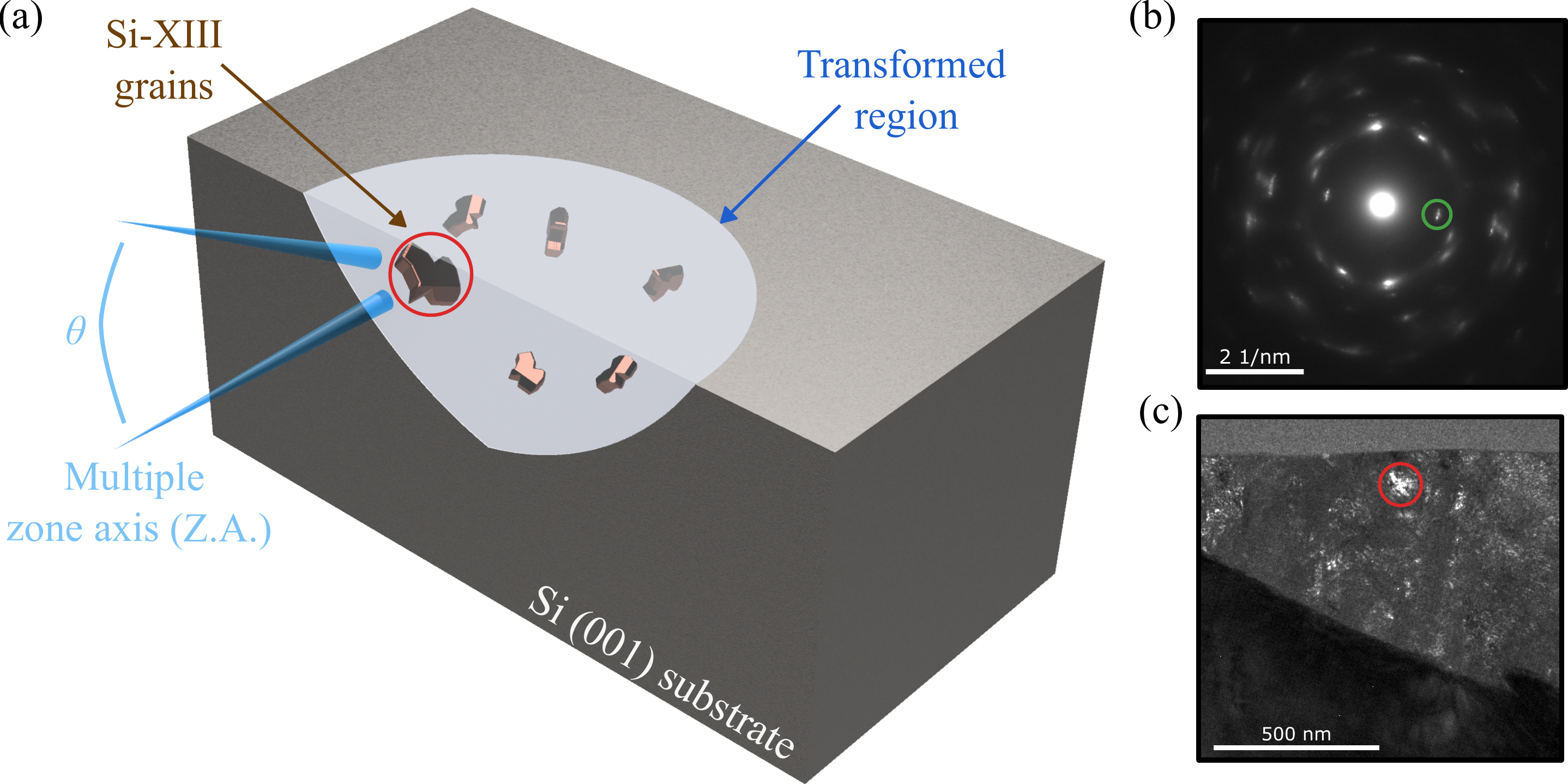}
    \caption{Identification of Si-XIII grains in the indented pit. (a) Schematic representation of the annealed pit: transformed region in gray and Si-XIII grains in brown. (b) SAED pattern of the transformed region in the pit. A spot (circled in green) is selected relative to the interatomic distances $4.3-4.6~\text{\AA}$. (c) Dark Field TEM obtained from the reflection marked by the green circle in (b).}
    \label{fig:grain}
\end{figure*}

The procedure began by selecting a prominent unknown diffraction spot visible in the SAED pattern from the pit's inner region, as shown in Figure~\ref{fig:grain}(a)-(c). Specifically, the spots positioned near the center of the pattern, indicated by a green circle in Figure~\ref{fig:grain}(b) and relative to the interatomic distances of $4.3-4.6~\text{\AA}$, have frequently been reported in the literature as an indication of the presence of Si-XIII ~\cite{Haberl2015, Wong2019_JAP_SiXIII}. An objective aperture was inserted to select this spot and generate a corresponding dark-field image, spatially locating the area of interest related to the unknown crystal structure, as reported in Figure~\ref{fig:grain}(c). 

\begin{figure*}[h!tb]
    \centering
    \includegraphics[width=01.0\linewidth]{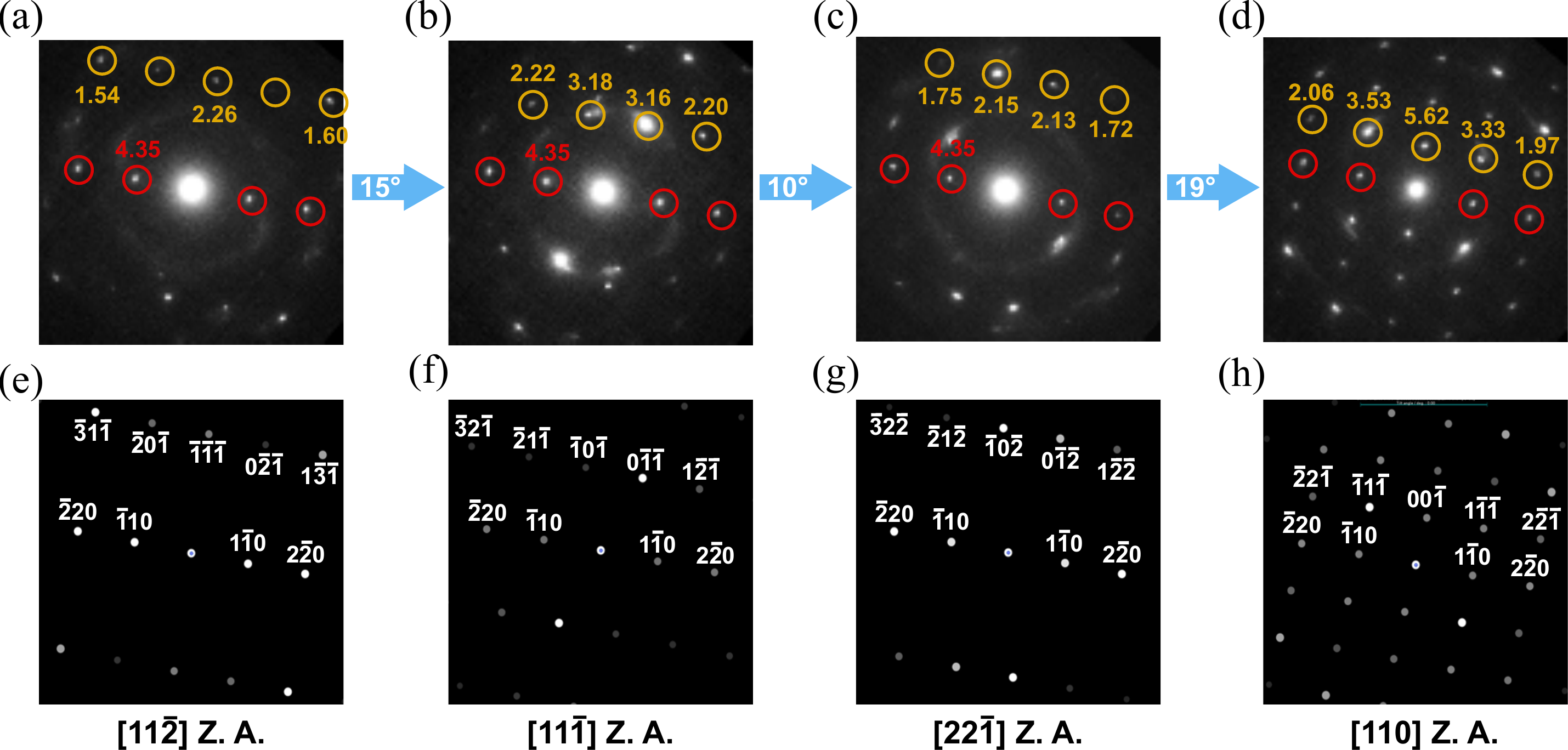}
    \caption{SAED analysis of the SI-XIII grain. One grain of the Si-XIII phase was selected (circled in red in Fig.~\ref{fig:grain}) and observed along multiple zone axis with mutual rotation of specific angles ($\theta$ in Fig.~\ref{fig:grain}). Panels (a)-(d) represent SAED patterns acquired from the Si-XIII grain across multiple zone axes with indicated mutual rotation angles. In these experimental patterns, the corresponding plane spacing distance is reported in \AA, for the main reflections. The sequence shows the systematic tilting procedure used to map the reciprocal lattice and the very good agreement with the proposed cell structure, as demonstrated by the corresponding simulated SAED (e)-(h), in which the pattern indexing is also reported.}
    \label{fig:saed}
\end{figure*}

From the dark field image, it is evident that the unknown phase is distributed throughout the entire pit area. The brighter grain in the dark field image (red circle in Figure~\ref{fig:grain}) was then selected, following its specific SAED response during the entire sample rotation procedure. The sample was then tilted around the axis parallel to the selected spot direction, maintaining it in Bragg condition until a low-index zone axis was reached. This orientation was set as the reference of the goniometer. Tilt series were then performed to span a total angle of $\sim$50$^\circ$ (goniometer limit), following the same grain and maintaining the same spot in Bragg condition, i.e. rotating around the direction perpendicular to the (1-10) planes, continuously tracking additional zone axes while recording precise rotation angles. This generated comprehensive rotational maps of the reciprocal lattice, from which compatible Bravais lattice parameters could be inferred.
A representative sequence of SAED patterns acquired from this grain is shown in Figure~\ref{fig:saed}(a)-(d). 

\begin{figure*}[htb]
    \centering
    \includegraphics[width=1.0\linewidth]{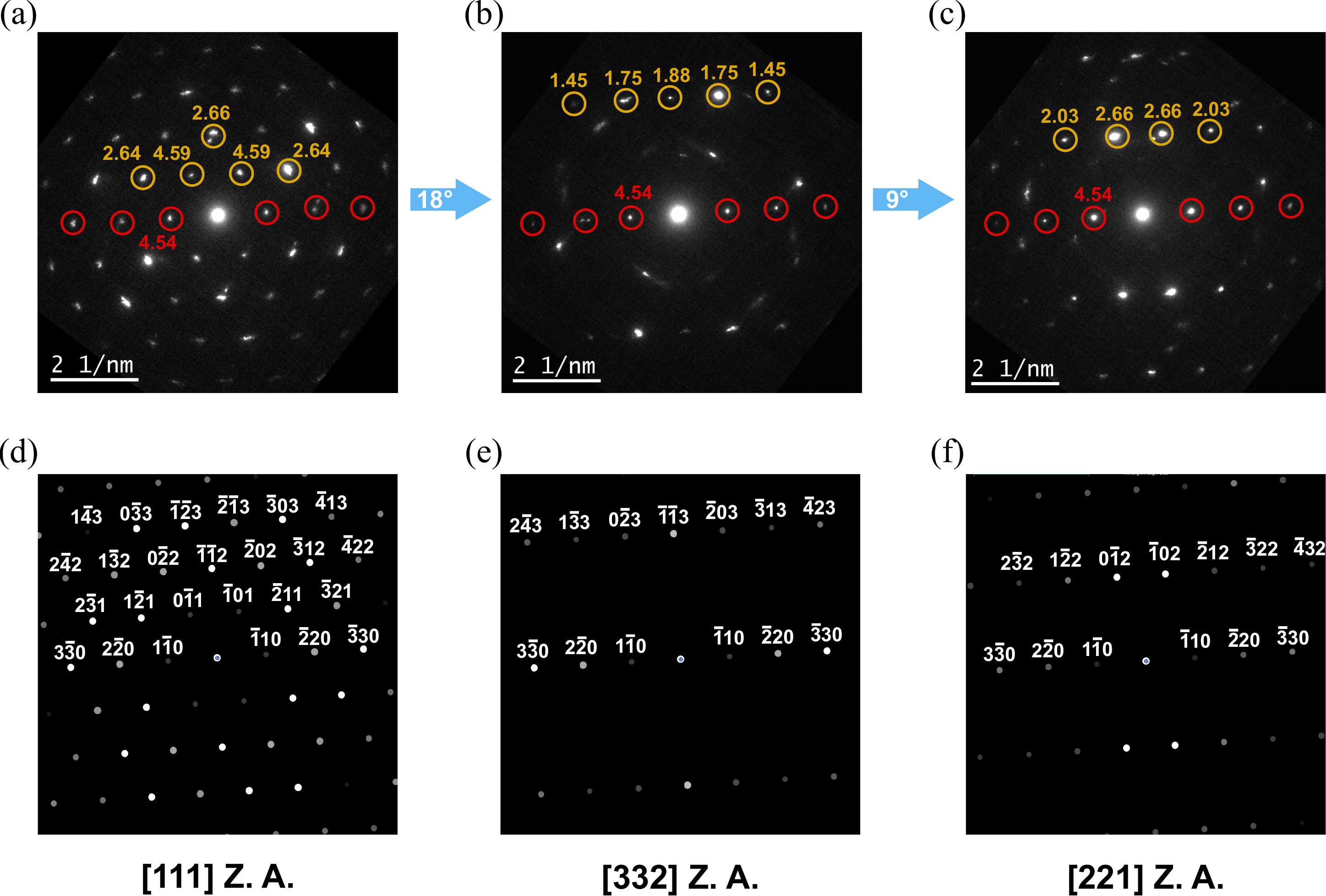}
    \caption{SAED identification of the matrix. (a)-(c) SAED patterns acquired from the same crystalline grain across multiple zone axes, with indicated mutual rotation angles. In these experimental patterns, the corresponding plane spacing distance is reported in \AA, for the main reflections. The sequence shows the systematic tilting procedure used to map the reciprocal lattice and a good agreement with the R8 cell structure, as demonstrated by the corresponding simulated SAED (d)-(f). Indexing of the R8 phase is presented in the rhombohedral basis, as defined in Table S2}
    \label{fig:saed2SI}
\end{figure*}

The zone axis morphologies and mutual angular relationships observed in the SAED patterns closely resemble those of the R8 structure, as evident from the comparison with the R8 SAED patterns shown in Figure~\ref{fig:saed2SI} (cf. FigureS1 in the Supporting Information, where the R8 simulated patterns are presented in the same zone axes). The R8 phase is indeed commonly formed during nanoindentation alongside BC8, as extensively discussed in the Raman spectroscopy section below and widely reported in the literature\cite{domnich2002phase, Ruffell2009, Haberl2015, Haberl2016_APR_ExoticSi, Wong2019_JAP_SiXIII, Bikerouin2025_SmallStruct_hdSi}. However, the interplanar spacings and unit cell angles derived from the SAED patterns, shown in Figure~\ref{fig:saed}(a)-(d), differ significantly from those of pristine R8. Guided by this structural similarity, we systematically explored distortions of the R8 crystal structure, varying lattice constants and angles to match the measured diffraction data. This iterative process ultimately yielded a set of cell parameters producing very good agreement between calculated and experimental interplanar distances, zone-axis geometries, and rotation angles. Simulated SAED patterns computed for this optimized cell, shown in Figure~\ref{fig:saed}(e)–(h) alongside their corresponding indexing, faithfully reproduce the experimental patterns in both zone-axis geometries and relative reflection intensities.

The procedure described above was also applied to investigate the crystalline matrix around the Si-XIII grains. Residual R8 crystals are clearly identified by the SAED analysis presented in Figure~\ref{fig:saed2SI}. Panels (a)-(c) display a series of diffraction patterns acquired from a single crystalline grain, recorded during a systematic tilting procedure across multiple zone axes. The mutual rotation angles and the measured d-spacings of the primary reflections (reported in \AA) exhibit high fidelity to the R8 silicon symmetry. A small distortion of the R8 matrix can be inferred from the d-spacings measurements. This is consistent with previous results reported on the R8 phase produced after indentation, as discussed in Ref.~\cite{wong2019formation}. The structural assignment is confirmed by the simulated SAED patterns in panels (d)-(f) which show excellent agreement with the experimental data (the crystallographic indexes are referred to the rhombohedral basis defined in Table S2). The systematic tilting procedure applied to this grain, spanning three zone axes with well-defined mutual rotation angles, confirms that some surrounding matrix retains the R8 crystal symmetry, ruling out any ambiguity in the structural assignment of the newly identified Si-XIII phase. The excellent match between experimental and simulated patterns thus provides an independent cross-check of the crystallographic framework used throughout this study.

Table S3 compares the measured and simulated d-spacings for each indexed (\emph{hkl}) reflection of the Si-XIII and R8 structures. The simulated values were obtained using the experimental lattice parameters reported in Table S2. For each reflection, the deviation between measured and simulated d-spacings is reported both as a value and as a percentage. The average of the absolute deviations was found to be 0.016 \AA\  for Si-XIII and 0.014 \AA for R8. These values are remarkably similar and lie within the estimated experimental uncertainty, which is generally 0.02 \AA or better for interplanar spacings below 4 \AA in a single crystal. This close agreement provides further support for the reliability of the proposed Si-XIII structural model and demonstrates a level of consistency between experiment and simulation comparable to that obtained for the well-established R8 phase.

Following a detailed reconstruction of the reciprocal lattice of Si-XIII crystals using SAED, we identified this phase as a distorted R8-like structure consistent with the experimental zone axes.  We then refined and validated candidate atomic structures using DFT.  Taking the R8 unit cell as an initial template, we systematically adjusted the lattice parameters and cell angles within the ranges allowed by the SAED constraints and, for each trial geometry, performed full variable-cell relaxations. For every relaxed configuration, the corresponding diffraction pattern was simulated and quantitatively compared with the experimental SAED data. Through this iterative procedure, we identified a structure whose simulated diffraction spots closely reproduce the measured zone-axis patterns and which, at the same time, lies at a local minimum of the DFT total energy, i.e., a thermodynamically metastable phase consistent with the Si-XIII assignment.

The relaxed Si-XIII structure consistently converges to a triclinic cell with space group $P\bar{1}$ (No.~2) for all exchange-correlation functionals considered (complete details are provided in the Methods section). The corresponding equilibrium lattice parameters obtained with LDA, PBEsol, PBE, and SCAN are summarized in Table~\ref{tab:cell_params}, together with the refined experimental values extracted from SAED. As expected, LDA yields the most compact cell, PBE the largest volume, while PBEsol and SCAN provide intermediate values. Starting from the DFT-relaxed cells, we re-simulated the SAED patterns and found that, in particular, the SCAN- and PBEsol-based structures reproduce the experimental diffraction spots with excellent accuracy. A subsequent fine-tuning of the lattice parameters to optimize the agreement with the measured SAED pattern led to the “Experimental” values reported in Table~\ref{tab:cell_params}. In all cases, the eight-atom unit cell is composed of silicon atoms occupying general Wyckoff $2i$ positions of $P\bar{1}$, with no additional symmetry-imposed constraints on the internal coordinates (full fractional coordinates are reported in the Supplementary Material file, Table S1).

Remarkably, the equilibrium density of this structure lies much closer to those of the cubic and hexagonal diamond phases than to the low-density R8 and BC8 metastable phases. Its DFT formation energy exceeds that of the stable diamond phases by $\Delta E \approx 100-120$~meV/atom but remains lower than R8 by $\Delta E \approx 20~-~90$~meV/atom (details are reported in the Supplementary Material file, Table S2), positioning Si-XIII as a competitive metastable allotrope, as also shown in Figure~\ref{fig:Raman10}(b). To the best of our knowledge, this crystal structure has not been reported previously in either theoretical predictions or experimental studies, and thus represents a genuinely new allotrope that naturally emerges as the resolution for the long-sought Si-XIII phase.

\begin{table}[t]
\centering
\begin{tabular}{lccccccc}
    \hline
    XC / data set & $a$ [\AA] & $b$ [\AA] & $c$ [\AA] & $\alpha$ [$^{\circ}$] & $\beta$ [$^{\circ}$] & $\gamma$ [$^{\circ}$] & $V$ [\AA$^{3}/\text{atom}$] \\
    \hline
    LDA                & 5.150 & 5.745 & 6.204 & 109.1 & 100.0 & 107.1 & 19.78 \\
    PBEsol             & 5.200 & 5.797 & 6.263 & 109.1 & 100.0 & 107.0 & 20.35 \\
    PBE                & 5.225 & 5.828 & 6.297 & 109.0 & 100.0 & 107.0 & 20.70 \\
    SCAN               & 5.191 & 5.807 & 6.266 & 109.1 & 100.0 & 107.0 & 20.38 \\
    Experimental       & 5.20  & 5.74  & 6.21  & 109   & 100   & 107   &  20.04 \\ 
    \hline
\end{tabular}
\caption{Equilibrium lattice parameters of the proposed Si-XIII structure obtained from DFT relaxations with different exchange-correlation functionals, together with refined experimental values from SAED. The space group is $P\bar{1}$ (No.~2) with 8 atoms per unit cell. Wyckoff Positions are provided in Table S1 of the Supplementary Material file.}
\label{tab:cell_params}
\end{table}

\subsection{Raman characterization and modeling}

To validate this structural assignment, we compare experimental and theoretically computed Raman spectra. Gogotsi and Domnich~\cite{domnich2002phase} first assigned the Si-XIII label to an unknown phase observed after nanoindentation at $150$--$200~^\circ$C, characterized by distinctive Raman peaks at $200$, $330$, $475$, and $497~\text{cm}^{-1}$ that could not be attributed to \emph{dc}, BC8, R8, or amorphous silicon. Subsequently, several other studies confirmed these Raman signatures for phases recovered after indentation-induced high-pressure transformations and subsequent annealing~\cite{Ruffell2009, Haberl2015, Haberl2016_APR_ExoticSi, Wong2019_JAP_SiXIII}.

However, a 2009 theoretical study~\cite{MylvaganamNanotech2009}, based on classical molecular dynamics simulations using the Tersoff potential (whose limitations in the present context are discussed in \cite{GeActaMat2024}), mislabeled a six-fold coordinated, high-density silicon phase observed during indentation simulations as Si-XIII. This misnaming has since been perpetuated in several theoretical works, despite the fact that the computationally predicted structure bears no relation to the experimentally observed Si-XIII phase recovered after annealing. The cell proposed here, after DFT optimization, finally reconciles the experimental observations with a well-defined crystal structure, as confirmed by the Raman validation discussed below.

\begin{figure*}[htb]
    \centering
    \includegraphics[width=1.0\linewidth]{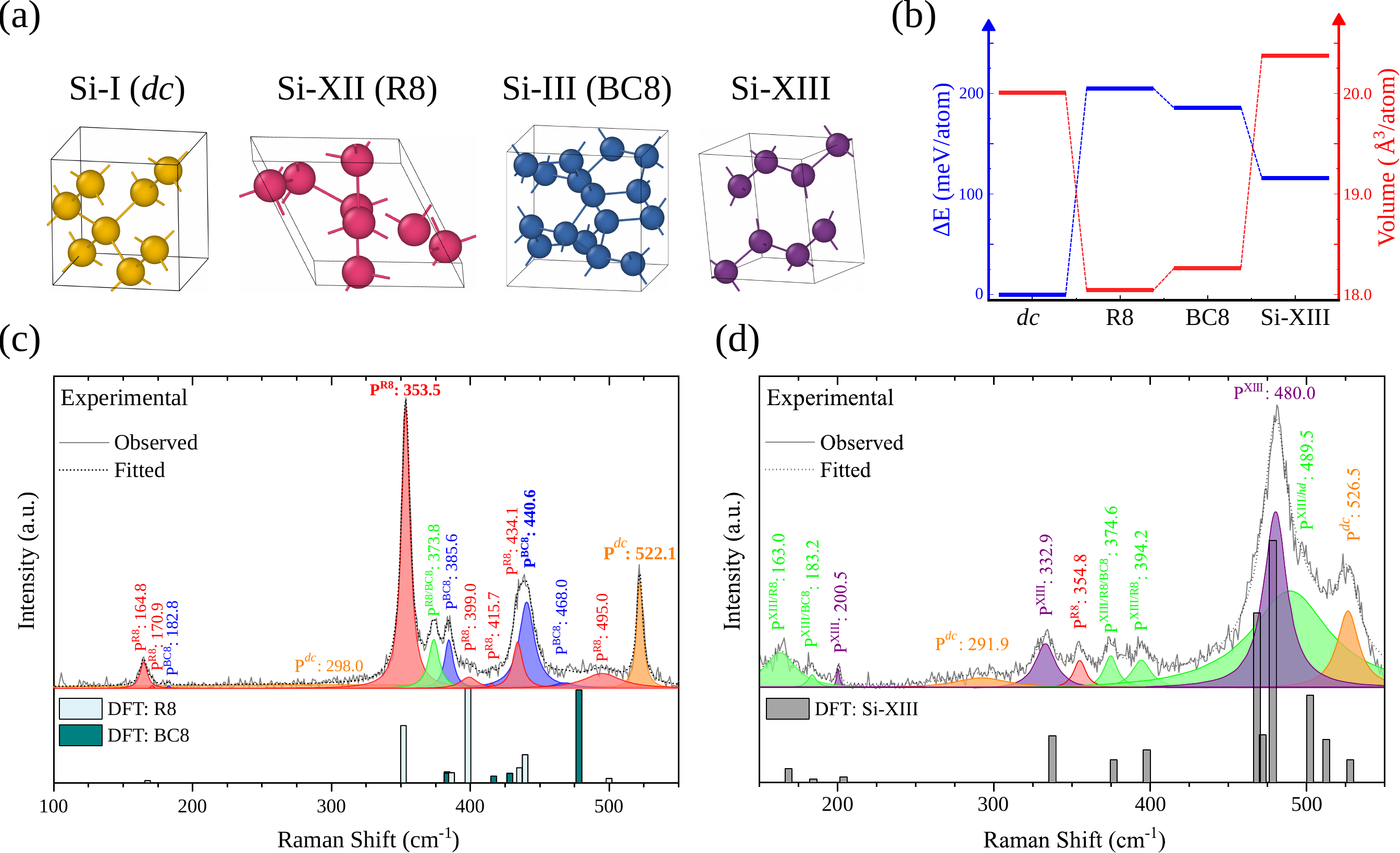}
    \caption{Raman spectroscopic analysis of silicon phases under nanoindentation and post-annealing. Panel (a) shows the the conventional cells of the analyzed phases: \emph{dc} (orange), R8 (magenta), BC8 (blue), and Si-XIII (purple). Their formation energies and volumes are reported in panel (b). Fitted experimental polarized Raman spectra acquired in a parallel polarization of 10~$\mu$m tip indented silicon are shown for (c) as-indented and (d) post-annealed samples. Peak positions, intensities, and widths were treated as free parameters in the fitting procedure. Raman-active mode frequencies for the different phases, as reported in Table~\ref{tab:Raman}, were used as references. Experimental peaks are represented by Lorentzian functions colored according to the respective crystal structures shown in panel (a). DFT-calculated non-resonant Raman frequencies, shown as vertical lines in panels (c) and (d), have been rigidly shifted upward by 12~cm$^{-1}$ to align the main R8 peak to the experimental one and facilitate comparison across all phases.}
    \label{fig:Raman10}
\end{figure*}

The Raman spectra of the sample in parallel polarization (see Methods section), both as-indented and post-annealed, are presented in Figure~\ref{fig:Raman10} and compared with theoretical spectra derived from first-principles calculations. All peak frequencies are listed in Table~\ref{tab:Raman}, together with previous experimental reference values. The as-indented spectrum of Figure~\ref{fig:Raman10}(c) is consistent with analogous spectra we reported in Ref.~\cite{Bikerouin2025_SmallStruct_hdSi}, where the presence of BC8, R8, and deformed \emph{dc} phases was unambiguously identified. Notice that in Ref.~\cite{Bikerouin2025_SmallStruct_hdSi} it is also shown that the scattering intensities of the Raman-active modes measured before annealing, which are related to phases different from \emph{dc}, are highest for parallel polarization, with the analyzer perpendicular to the indentation direction. This is confirmed in the current study and discussed further below.  Therefore, we first focus on the data obtained in parallel polarization. This helps identify both R8 and BC8 although they appear as a mixture after nanoindentation as described in \cite{Bikerouin2025_SmallStruct_hdSi} and still exhibit several closely spaced peaks.  

The peak frequencies obtained after fitting the as-indented spectra (Table~\ref{tab:Raman}) are remarkably consistent with both previously reported experimental values~\cite{johnson2011temperature, Ruffell2009} and calculated ones.  While the calculated frequencies are slightly shifted due to the typical underestimation of Raman frequencies in LDA-DFT calculations~\cite{Bikerouin2025_SmallStruct_hdSi}, this is corrected in the theoretical peaks shown in Figure~\ref{fig:Raman10}.  This correction allows for a clear assignment of each peak to either BC8 or R8 (indicated by bold text in Table~\ref{tab:Raman}), except for the peak at approximately 374 cm$^{-1}$, which is shared between both phases.

Most of these Raman peaks disappear after thermal annealing, as shown in Figure~\ref{fig:Raman10}(d), indicating a strong phase transition. In particular, the two most intense peaks of R8 and BC8, found at approximately 354 and 441~cm$^{-1}$, respectively, exhibit minimal intensity (the former) or are completely absent (the latter). In fact, only a few Raman modes previously attributed to the R8/BC8 phases in Table~\ref{tab:Raman} remain detectable and have marginal intensities. This suggests that some residual R8/BC8 phase is still present, but with the majority of this mixture transformed into a crystallographically distinct structure. 

Remarkably, the Raman mode shifts calculated for the proposed Si-XIII structure, shown together with the experimental spectra in Figure~\ref{fig:Raman10}(d), perfectly match those corresponding to the peaks appearing only after the annealing process. Moreover, Table~\ref{tab:Raman} shows that the frequencies of these theoretically predicted modes are in excellent agreement with previous reports on the, not identified yet, Si-XIII phase. Specifically, we find experimental peaks at approximately 200, 333, 480, and 490~cm$^{-1}$, all of which correspond to Raman modes calculated using DFT for our proposed Si-XIII structure. Noteworthy, these Raman mode shifts are in nearly perfect match with those reported in previous experiments at  200, 330, 475, and 497~cm$^{-1}$, respectively~\cite{domnich2002phase}. Although the peak at $\sim 480~\text{cm}^{-1}$ is the most distinctive Raman fingerprint of the Si-XIII phase in our spectra, the peaks at 200 and 330~cm$^{-1}$,  not present before annealing, can also unambiguously identify Si-XIII based on the DFT calculations. 

Additionally, there are additional peaks at $\sim163$, $183$, $375$, $394$, and $490~\text{cm}^{-1}$ that, based on our calculations, could also be attributed to Si-XIII, but are shared also with the BC8/R8 phases. Moreover, it is often difficult to identify single peaks and assign them to a specific phase, particularly in the $490-520~\text{cm}^{-1}$ range. In fact, in this range, in addition to several possible peaks attributable to BC8/R8 or \emph{dc}, a couple of peaks attributable to \emph{hd} are also present~\cite{Bikerouin2025_SmallStruct_hdSi}. Indeed, according to previous reports~\cite{Haberl2015, Wong2019_JAP_SiXIII}, \emph{hd} is likely present in our samples after thermal annealing at $220^\circ~\text{C}$. Consequently, the shoulder at approximately 500~cm$^{-1}$ in our spectra, giving rise to the Lorentzian function at approximately 490~cm$^{-1}$ shown in Figure~\ref{fig:Raman10}d, may be attributed to a minimal fraction of \emph{hd}.

\begin{table*}[!t]
\centering
\small
\setlength{\tabcolsep}{4pt}
\begin{tabular*}{\textwidth}{@{\extracolsep{\fill}} lcccccc @{}}
\toprule
\textbf{Phase} & \textbf{Raman} & \multicolumn{3}{c}{\textbf{This work}} & \textbf{Literature} \\
\cmidrule(lr){3-5} 
& \textbf{Mode} & \textbf{Exp. before ann.} & \textbf{Exp. after ann.} & \textbf{Theo.} & \textbf{Exp.\cite{johnson2011temperature}}\\
\midrule
\emph{dc} & T$_{2g}$ & \textbf{522.1} & \textbf{526.5} & 513.4 & 520.3 \\
\midrule
\multirow{8}{*}{R8} & A$_g$ & 164.8 & 163.0 & 155.9 & 164.8 \\
& E$_g$ & \textbf{170.9} & - & 163.0 & 170.0 \\
& A$_g$ & \textbf{353.5} & \textbf{354.8} & 340.0 & 351.9 \\
& E$_g$ & 373.8 & 374.6 & 374.6 & 373.3 \\
& A$_g$ & \textbf{399.0} & 394.2 & 386.4 & 397.1 \\
& A$_g$ & \textbf{415.7} & - & 423.4 & 412 \\
& E$_g$ & \textbf{434.1} & - & 427.5 & - \\
& E$_g$ & \textbf{495.0} & 489.5 & 487.9 & - \\
\midrule
\multirow{5}{*}{BC8} & T$_g$ & \textbf{182.8} & 183.2 & 155.1 & 182.4 \\
& T$_g$ & 373.8 & 374.6 & 371.2 & 373.3 \\
& A$_g$ & \textbf{385.6} & - & 371.3 & 384.2 \\
& T$_g$ & \textbf{440.6} & - & 416.5 & 437.5 \\
& E$_g$ & \textbf{468.0} & - & 466.2 & 463 \\
\midrule
\multirow{3}{*}{Si-XIII} & A$_{g}$ & 165.0 & 163 & 156.5 & - \\
& A$_{g}$ & - & 183.2 & 172.4 & - \\
& A$_{g}$ & - & \textbf{200.5} & 191.8 & 200 \\
& A$_{g}$ & - & \textbf{332.9} & 325.5 & 330 \\
& A$_{g}$ & 374.0 & 374.6 & 364.6 & - \\
& A$_{g}$ & - & 394.2 & 385.7 & - \\
& A$_{g}$ & - & - & 456.3 & - \\
& A$_{g}$ & - & - & 459.9 & - \\
& A$_{g}$ & - & \textbf{480.0} & 466.3 & 475 \\
& A$_{g}$ & - & 489.5 & 490.3 & 497 \\
& A$_{g}$ & - & - & 500.5 & - \\
& A$_{g}$ & - & - & 515.9 & - \\
\bottomrule
\end{tabular*}
\caption{Comparison of experimental and theoretical Raman frequencies (in cm$^{-1}$) for various investigated silicon phases. The irreducible representations of the experimental phonon modes have been assigned through correspondence with the calculated modes. Previously reported experimental values for \emph{dc}, R8, and BC8~\cite{johnson2011temperature}, as well as Si-XIII~\cite{domnich2002phase}, are included for reference and are representative of the frequencies reported in other studies~\cite{wong2019formation, smillie2020exotic, mannepalli2019situ, Wong2019_JAP_SiXIII, Ge2004, Ruffell2009}. Experimental values from this work are highlighted in bold when the assignment is unique to a certain phase.}
\label{tab:Raman}
\end{table*}

Furthermore, the peak located at approximately $522~\text{cm}^{-1}$ ($526~\text{cm}^{-1}$) is present before (after) annealing. This peak is certainly attributable to strained \emph{dc} not transformed into the metastable phases. However, the fact that its shift from the pristine phase peak expected at approximately 520~cm$^{-1}$ increases after annealing suggests an increasing compressive strain acting on \emph{dc}, likely due to the expansion of the metastable phases upon annealing. This confirms the volume increase expected from the transition from the R8/BC8 mixture to the Si-XIII phase, in agreement with the calculated volumes reported in Table~\ref{tab:cell_params} and further discussed below.

\begin{figure*}[tbh]
\centering
	\includegraphics[width=1.0\textwidth]{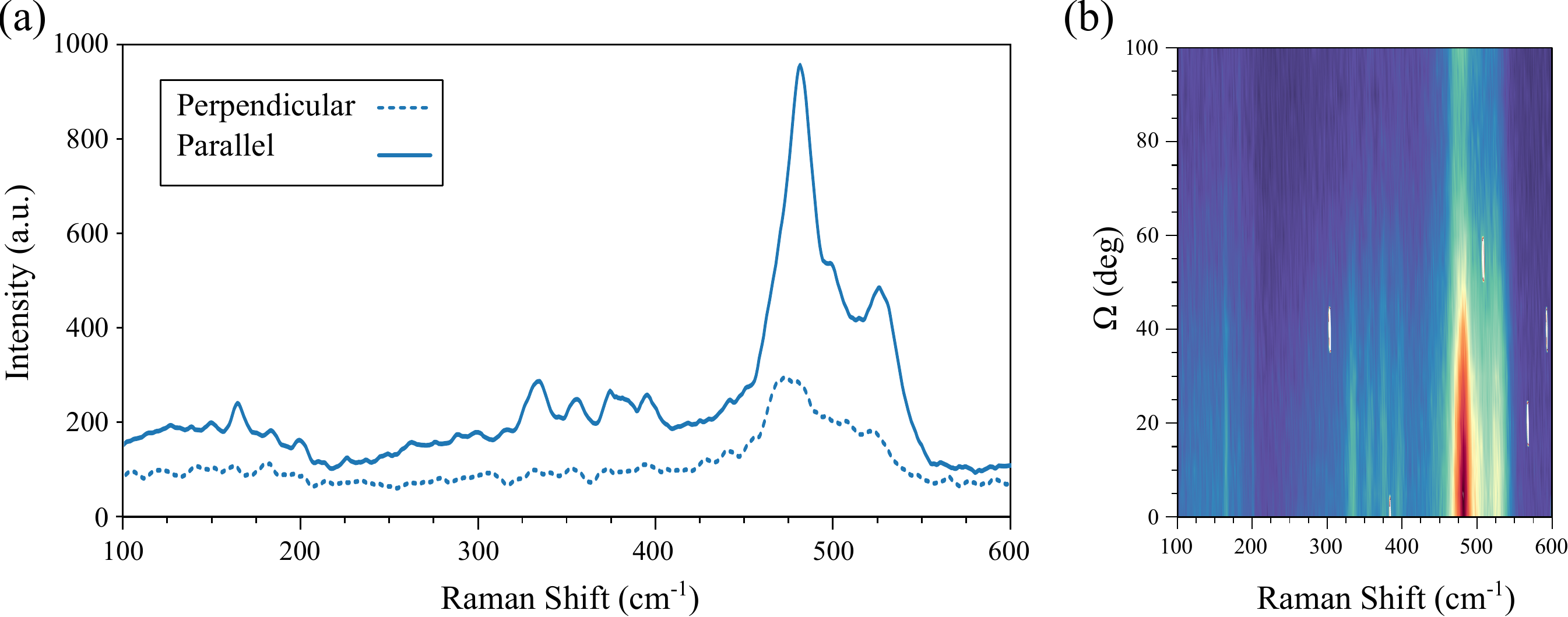}
	\caption{Polarized Raman analysis for indented silicon after annealing. In panel (a) both perdendicular (dashed line) and parallel (solid line) configurations are shown. The full polarized spectrum taken after the annealing procedure and as a function of incidence polarization angle ($\Omega$) is shown in (b).}
	\label{fig:raman_pol}
\end{figure*}

Finally, Figure~\ref{fig:raman_pol} presents the polarisation-dependent vibrational spectra after annealing. Panel (a) shows the spectra acquired in perpendicular (dashed line) and parallel (solid line) polarization (see Methods section for details about the Raman configuration). Furthermore, the full map of the vibrational modes as function of the incidence polarization angle ($\Omega$) was acquired and the resulting spectra is displayed in Panel (b). The angular evolution of the Raman peaks reveals a decreasing of the peak intensities from the parallel to the perpendicular, leading to almost the disappearance of the low-frequency ones, as in the case of BC8/R8 discussed above. This polarization dependence mirrors the behavior previously observed for the R8 and BC8 phases~\cite{Bikerouin2025_SmallStruct_hdSi}, suggesting that the Raman-active modes of Si-XIII share a similar orientational response, consistent with the structural kinship between these phases established in Section~\ref{sec:res_SAED}. The full angular map in panel (b) thus provides a compact, two-dimensional fingerprint of the Si-XIII phase that can serve as a reference for its identification in future experimental studies.

\subsection{Kinetics of the Si-XIII phase}

The kinetics of the Si-XIII phase formation is then investigated to establish its connection to the main Si phases involved in these process conditions: \emph{dc}, \emph{hd}, R8, and BC8. Using extensive PES explorations by means of the SS-Dimer method (see Methods section),  we map the kinetic barriers connecting these phases, specifically focusing on the Si-XIII candidate structure.

\begin{figure*}[!b]
\centering
	\includegraphics[width=0.9\textwidth]{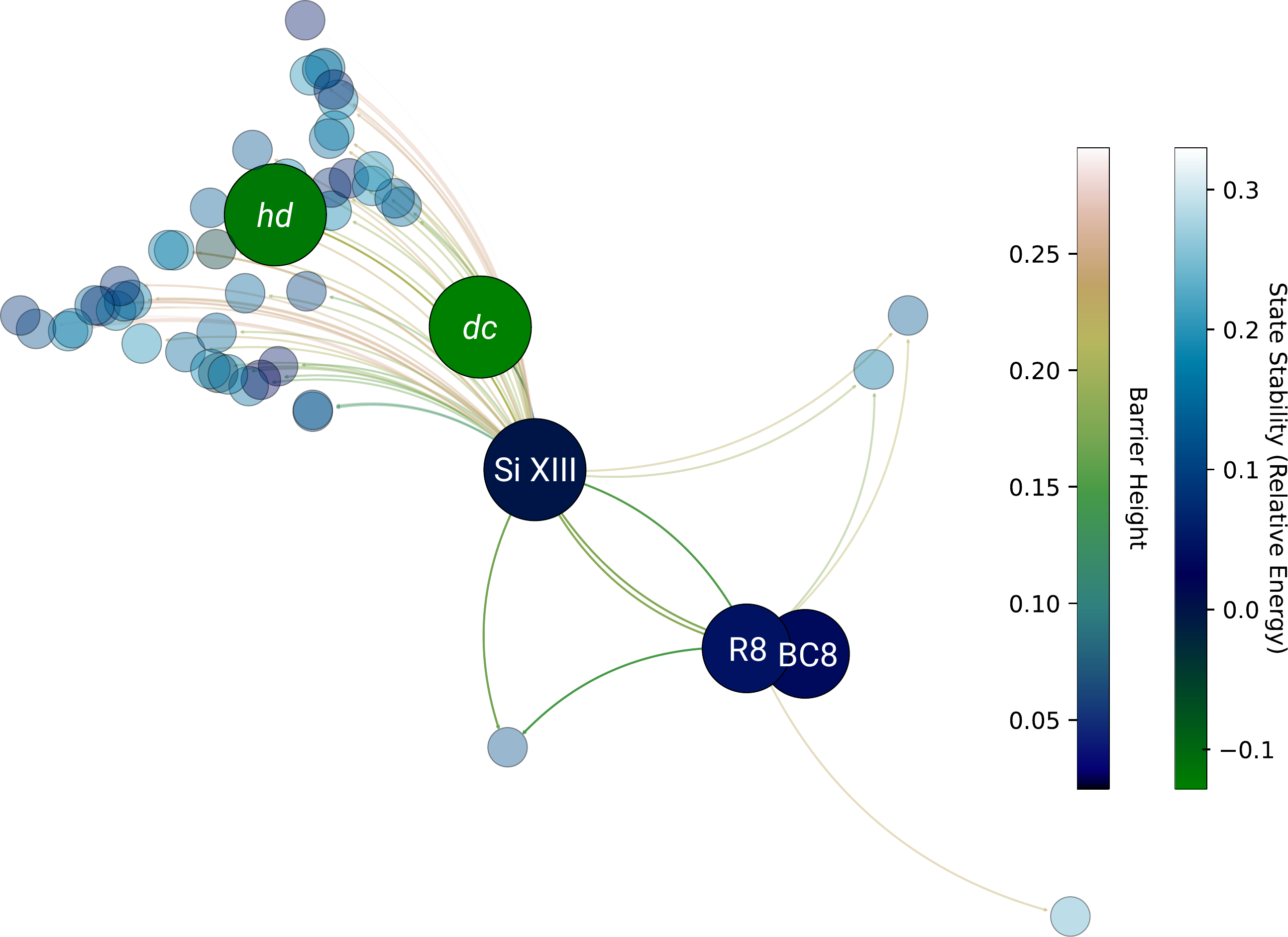}
	\caption{Transition Pathway Network showing the interconnection between the Si-XIII phase and the main known phases of silicon. The nodes of the network represent metastable minima while the edges connecting them represent the transition barrier found by the SS-Dimer method. The two colorbars represent the formation energy $\Delta E$ relative to the Si-XIII phase and the Barrier height for the transition connecting the two nodes. The relative length of the connection is proportional to the energy barriers of the transitions following the Kamada-Kawai representation.}
	\label{fig:network}
\end{figure*} 

Previous computational studies have attempted to find the Si-XIII phase or identify kinetic pathways for the~\emph{dc} phase transformations, exploiting a variety of approaches ranging from stochastic PES sampling to more sophisticated methods~\cite{Zhu2015,Zhu2019,Mujica2015,Yao2012,Behler2008,Zhao2012}. However, the exceptionally rich local-minima landscape of silicon makes these searches challenging: starting from \emph{dc} or R8/BC8, one typically encounters a multitude of competing minima without naturally landing on a phase that simultaneously bridges the high-pressure and diamond-like PES neighborhoods. Here, initiating the search directly from Si-XIII reveals its pivotal role as the missing connecting phase between these energy basins, as clearly illustrated in the transition pathway network of Fig.~\ref{fig:network}.

The SS-Dimer search was performed starting from randomly displaced configurations of the Si-XIII candidate structure, with around $20000$ repeated iterations. The details on the dimer iterations are provided in the Method section. The results were then analyzed and ordered by the height of the energy barrier to highlight the most relevant transitions. The same procedure was repeated with the other four phases under consideration: \emph{dc}, \emph{hd}, R8, and BC8. The final results are summarized in the Transition Pathway Network of Figure~\ref{fig:network}, where circles represent metastable minima (with the most relevant ones highlighted by larger circles and the others shaded), and connecting lines represent the kinetic barriers. The data in the plot are arranged following the Kamada-Kawai algorithm~\cite{KamadaIPL1989}, with line lengths proportional to the kinetic barrier heights between the connected minima. Transitions characterized by kinetic barriers higher than $0.3~\text{eV/atom}$ are not shown. Two color bars are reported for the height of the kinetic barrier and the formation energy of the energy minima, respectively.

\begin{figure*}[ht]
\centering
	\includegraphics[width=1.0\textwidth]{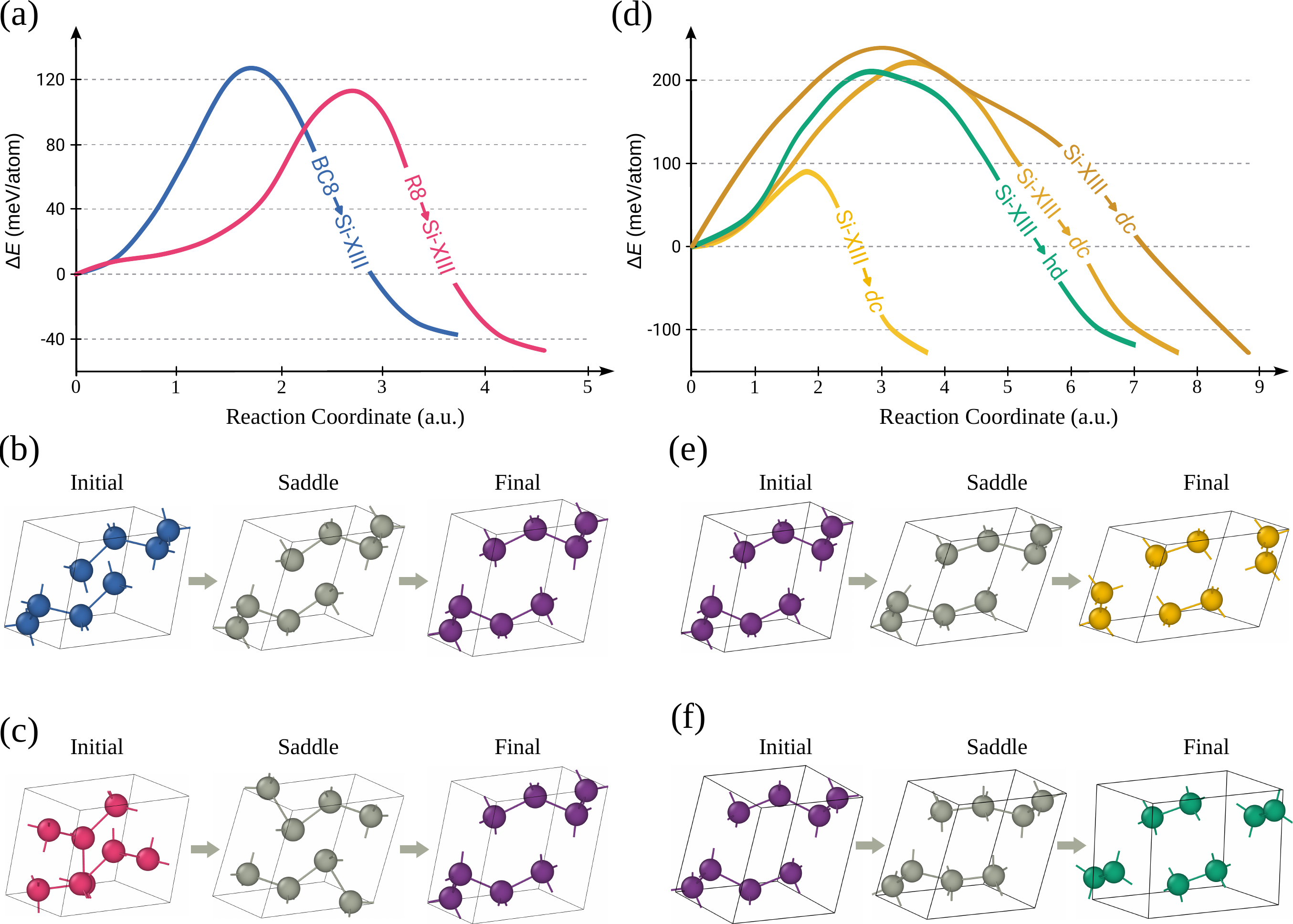}
	\caption{Minimum Energy Paths as calculated by Solid State Nudged Elastic Band connecting the Si-XIII phase to the main metastable phases of silicon. (a) Illustrates energy paths connecting the BC8 and R8 phases to the Si-XIII, while (d) shows paths connecting the Si-XIII phase to the \emph{dc} and \emph{hd} phases.  Atomic structures of the initial, saddle, and final states for the lowest energy barrier pathways illustrated in (a) and (d) are shown in (b)-(c) and (e)-(f), respectively.}
	\label{fig:NEB}
\end{figure*}

The central role of Si-XIII in the kinetics of silicon phase transitions is evident from Figure~\ref{fig:network}: it acts as a connecting phase between the metastable R8/BC8 phases and the stable \emph{dc} configuration. In fact, one result of our statistical investigation is that the transition path towards the Si-XIII phase represents the lowest kinetic pathway escaping from the energy basin of the R8 metastable phases. Additionally, we found a similar kinetic pathway that escapes from the BC8 phase and ends in the Si-XIII phase. The energy barriers associated with these transitions are: $111~\text{meV/atom}$ for the path connecting the R8 phase to the Si-XIII and $126~\text{meV/atom}$ for the one between the BC8 and the Si-XIII. These minimum energy paths are reported in Figure~\ref{fig:NEB}(a) and the atomic structures corresponding to their initial, final, and saddle-point are illustrated in Figure~\ref{fig:NEB}(b)-(c). These findings confirm the close relationship between the Si-XIII crystal structure and the R8 metastable phase, as already discussed from the crystallographic point of view in Section~\ref{sec:res_SAED} and in Supplementary Figure S1.

Another interesting observation coming from our analysis is that the Si-XIII phase is very closely connected to the stable \emph{dc} phase. We found three different transition pathways connecting these two phases (within our energy threshold for saddle point location), the lowest energy barrier carrying a barrier of only $90~\text{meV/atom}$, significantly lower the other two higher paths ($222$ and $243~\text{meV/atom}$), all reported in Figure~\ref{fig:NEB}(d) with the corresponding atomic structures in Figure~\ref{fig:NEB}(e)-(f).  Additionally, we found an energy path connecting the Si-XIII and the \emph{hd} phases. The energy barrier associated with this transition is $209~\text{meV/atom}$. This indicates that the direct evolution toward the \emph{dc} phase transition is kinetically accessible and favored within the present potential energy surface model. However, this result does not exclude \emph{hd} as an intermediate under experimental annealing conditions, where residual stress, defects, grain boundaries, phase coexistence, and thermal-history effects may alter the relative accessibility of competing pathways, as we have explicitly shown in Ref.~\cite{RovarisMTN2025} for the BC8 to \emph{hd} phase transition under residual stress. Taken together, these results embed Si-XIII into a coherent kinetic framework, confirming a physically motivated argument for the proposed crystalline structure.

Furthermore, as already mentioned, the crystalline structure of the Si-XIII phase does not show any high-level symmetry, being associated to the triclinic (P$\overline{1}$) space group. This is a peculiar property for a silicon allotrope, given that the lowest-energy fourfold-coordinated state (diamond-cubic) is highly symmetric. However, this observation can be made directly from our analysis of Si landscape. As we have shown, Si-XIII lies in a region of the configurational space that sits in between the low-energy diamond phases (\emph{dc} and \emph{hd}) and the higher energy BC8/R8. We thus performed a structural analysis of all these phases to understand the rearrangement of atomic bonds in terms of bond lengths and angles and their standard deviations, reported in Table~\ref{tab:bonds}. As shown in the Table, the bond lengths remain close to the diamond-cubic value for all the considered phases. Nevertheless, Si-XIII exhibits the broadest distribution of individual nearest-neighbor bond lengths, as quantified by both $\sigma_d$ and the $d_{\min}$--$d_{\max}$ interval reported in Table \ref{tab:bonds}. This local bond-length inhomogeneity is consistent with its triclinic $P\bar{1}$ symmetry, which does not impose equivalence among the nearest-neighbor bonds. However, the Si-XIII phase has already essentially recovered the cell volume of diamond-cubic silicon, whereas the R8/BC8 precursors are substantially denser ($\approx 11-12\%$ smaller volume than \emph{dc}). Despite this volume recovery, the bond-angle standard deviation of Si-XIII ($\sigma_{\alpha} \approx 9.7^{\circ}$) remains comparable to that of BC8 and far from the essentially zero value of \emph{dc}, indicating that the angular disorder has not relaxed altogether with the atomic volume. This is consistent with density changes in fourfold-coordinated silicon being accommodated primarily through bond-angle rather than bond-length distortion as already discussed in Ref~\cite{PiltzPRB1995}.

\begin{table}[h]
\centering
\begin{tabular}{llcccccc}
\toprule
    Phase& $d_{\text{mean}}$ [\AA]& $d_{\text{min}}$ [\AA] & $d_{\text{max}}$ [\AA] & $\sigma_d$ [\AA] & $\alpha_{\text{mean}}$[\textdegree] & $\sigma_{\alpha}$[\textdegree] & V[\AA$^3$]/\text{atom}\\
\midrule
    \emph{dc}   & 2.3528 & 2.3528 & 2.3528 & 0.00     & 109.47    & 0.00      & 20.05 \\
    R8         & 2.3669 & 2.3589 & 2.4173  & 0.0139   & 107.68    & 13.12      & 17.73 \\
    BC8          & 2.3644 & 2.3350 & 2.3742 & 0.0170   & 108.17    & 9.69     & 18.00 \\
    Si-XIII     & 2.3564 & 2.3109 & 2.3948 & 0.0242   & 109.25    & 9.75      & 20.35 \\
\bottomrule
\end{tabular}
\caption{Mean, minimum and maximum bond length ($d_{\text{mean}}$, $d_{\text{min}}$ and $d_{\text{max}}$) and its standard deviation $\sigma_d$, mean bond angle $\alpha_{\text{mean}}$ and its standard deviation $\sigma_\alpha$, and atomic volume $V$ per atom for the \emph{dc}, BC8, R8, and Si-XIII phases of silicon. Based on the PBEsol calculations reported in Table S2, bond lengths and angles were computed from the four nearest neighbors of each atom ($d_{\text{mean}}$, $\sigma_d$) and from the six angles subtended by pairs of these neighbors ($\alpha_{\text{mean}}$, $\sigma_\alpha$), averaged over all atoms in the unit cell. $V$ is the cell volume divided by the number of atoms.}
\label{tab:bonds}
\end{table}

We thus speculatively propose that there are at least two kinetically distinct pathways connecting dense, disordered high-pressure precursors to the low-pressure, high-symmetry phases: (i) a fast pathway in which volume recovery outpaces angular reordering and produces a low-symmetry, volume-relaxed but angularly disordered intermediate phase, Si-XIII;  (ii) a slower pathway in which volume and angular order relax more cooperatively — likely starting from the already comparatively more ordered BC8 precursor ($\sigma_{\alpha} \approx 9.7^{\circ}$ compared with $\approx 13.1^{\circ}$ in R8) — thereby bypassing a triclinic intermediate and connecting more directly to hexagonal-diamond or diamond-cubic Si. Within this framework, the triclinic symmetry of Si-XIII rather reflects the presence of a transient, symmetry-broken state along the possible transformation pathways.

\begin{figure*}[!t]
\centering
	\includegraphics[width=1.0\textwidth]{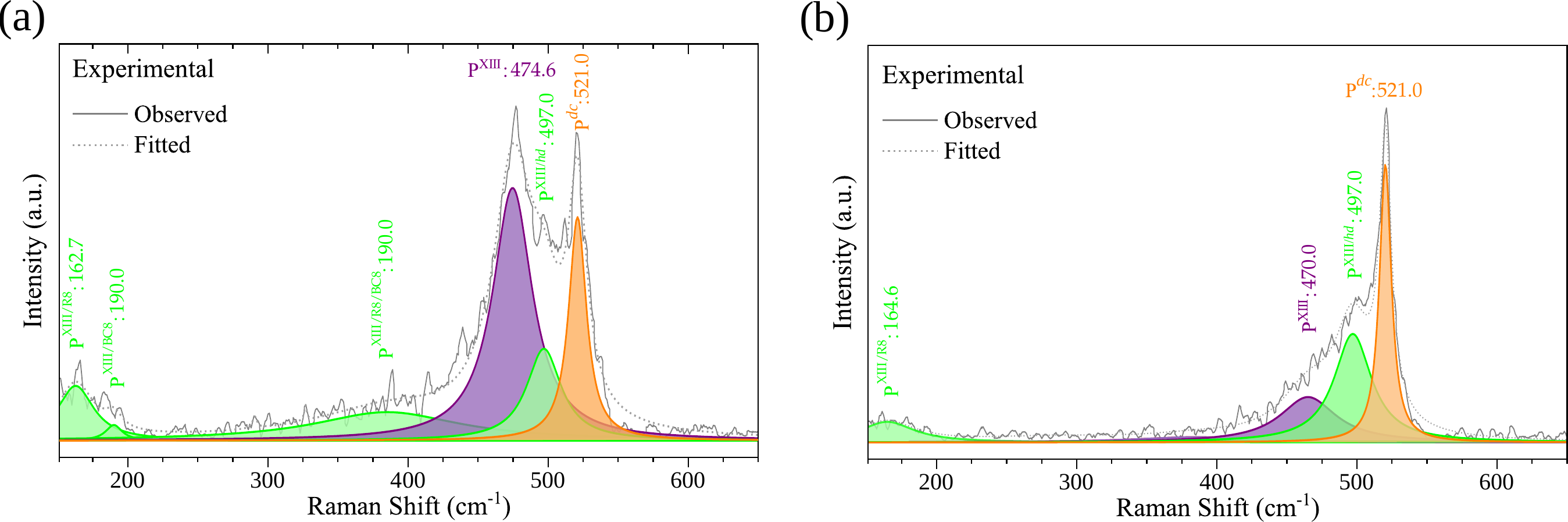}
	\caption{Analysis of the evolution of the Si-XIII phase before and after a second oven annealing. Fitted experimental polarized Raman spectra in parallel scattering geometry of 20~$\mu$m tip indented silicon are shown for a sample after a first stage annealing up to $220^{\circ}\text{C}$ (a) and the same sample after a successive oven annealing up to $250^{\circ}\text{C}$ (b). Peak positions, intensities, and widths were treated as free parameters in the fitting procedure. Raman-active mode frequencies for the different phases, as reported in Table~\ref{tab:Raman} are used as references. Fitted peaks are represented by Lorentzian functions colored according to the respective crystal structures as in Figure~\ref{fig:Raman10}.}
	\label{fig:Postannealing}
\end{figure*}

In order to validate these findings, in Figure~\ref{fig:Postannealing}, we report a Raman analysis of the post annealing evolution of the Si-XIII phase. First, the annealing procedure described in the Methods Section was reproduced on a silicon sample indented with a $20~\mu\text{m}$ spherical tip to produce the Si-XIII phase. We annealed the sample employing a heated stage holder up to $220^{\circ}\text{C}$, obtaining the result reported in Fig~\ref{fig:Postannealing}(a). The formation of the Si-XIII phase is clearly indicated by the Raman peak around the frequency of $480~\text{cm}^{-1}$, and some minor ones at about $160~\text{cm}^{-1}$ and $200~\text{cm}^{-1}$. In addition, a peak at $520~\text{cm}^{-1}$ clearly indicates the presence of \emph{dc}, while a lesser intense peak around $500~\text{cm}^{-1}$ may indicate some presence of the \emph{hd} phase~\cite{Bikerouin2025_SmallStruct_hdSi}. Overall, these results are perfectly consistent with the ones reported in Figure~\ref{fig:Raman10}(d), obtained with the $10~\mu\text{m}$ spherical tip. We then used a second annealing procedure on this sample by placing it in a furnace at $250^{\circ}\text{C}$. The Raman analysis after this second annealing step is reported in Figure~\ref{fig:Postannealing}(b). The analysis shows a clear reduction in the peaks associated with the Si-XIII phase after this second annealing process. Only a broad peak around the frequency of 520 cm$^{-1}$ with a long tail towards lower frequencies remains after the second annealing. This peak is mainly attributed to the \emph{dc} phase, with its long tail possibly indicating the presence of some \emph{hd} phase. This additional experiment directly proves that the Si-XIII phase is not stable at $250^{\circ}\text{C}$ and transforms to more stable phases, with its main reaction outputs being the \emph{dc} and possibly the \emph{hd} phases ~\cite{Wong2019_JAP_SiXIII,Ge2004}. The reproducibility of this behavior across two different indenter tip radii, 10~$\mu$m and 20~$\mu$m, confirms that the Si-XIII phase is not an artifact of a specific loading geometry, but a robust product of the R8-to-Si-XIII kinetic pathway under the employed thermal annealing conditions.

\begin{figure*}[!t]
\centering
	\includegraphics[width=1.0\textwidth]{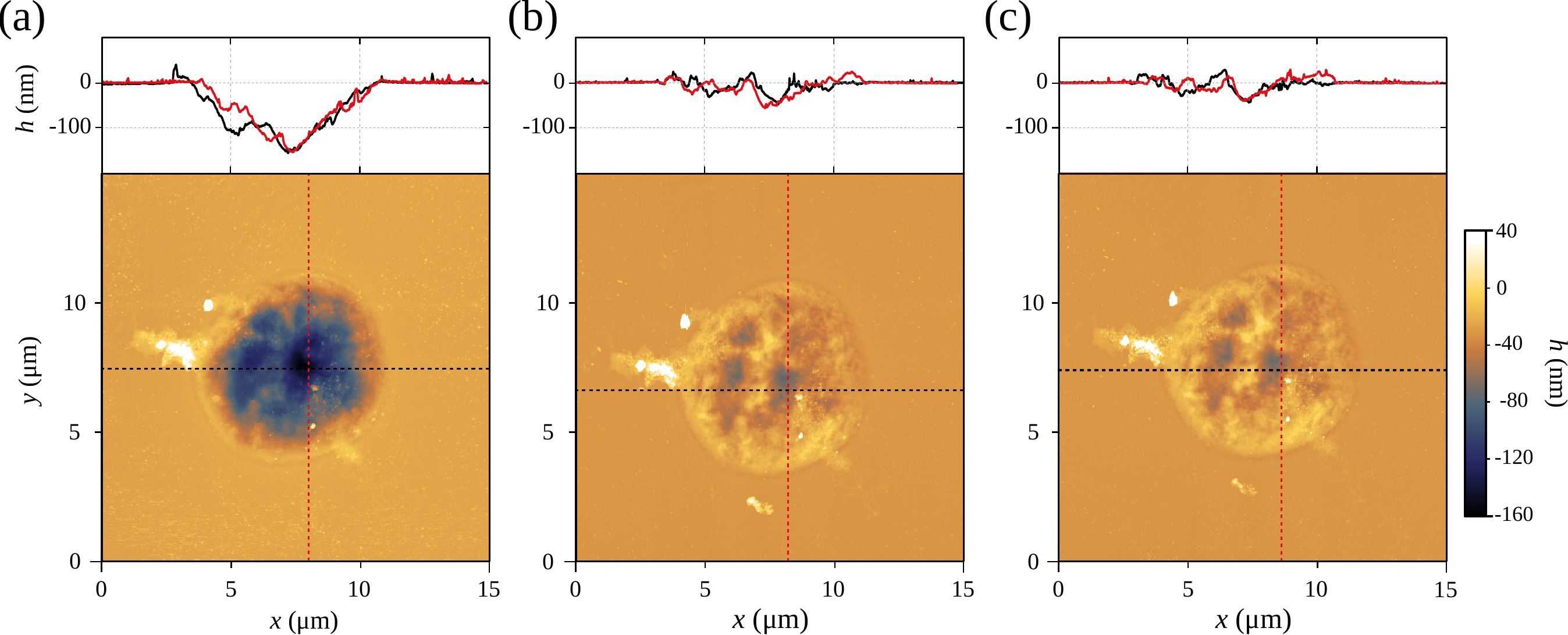}
	\caption{Atomic Force Microscopy analysis of the indented pit. The measurements are are reported as an AFM maps and a plot of the line scan along vertical and horizontal directions for a specific pit and are taken right after the indentation (a), after the first stage annealing (b) and after the final oven annealing (c).}
	\label{fig:AFM}
\end{figure*}

Additionally, Atomic Force Microscopy measurements were performed on the indented pit right after the indentation (see Figure~\ref{fig:AFM}(a)), then after the first stage annealing (Figure~\ref{fig:AFM}(b)) and after the final oven annealing (Figure~\ref{fig:AFM}(c)), and are shown together with the corresponding line profiles along two orthogonal directions. 
The results reveal significant differences in the indentation imprint heights after the first stage annealing, as compared to the as-indented pit. This is due to the volume per atom of Si-XIII being much larger than that of the R8/BC8 phases. However, almost no change is observed in the imprint height after the second annealing at 250$^\circ$C, confirming the volume similarity between Si-XIII and the Si diamond phases, as discussed previously. The quantitative line profiles allow a precise 
tracking of the volume changes at each annealing step, and the observed surface topography evolution is in agreement with the DFT-predicted volume sequence $V_\mathrm{R8/BC8} < V_\mathrm{Si\text{-}XIII} \approx V_\mathrm{dc}$, providing a direct macroscopic confirmation of 
the proposed structural assignment independently of any diffraction or spectroscopic measurement.

These results perfectly validate the kinetic modeling described above, demonstrating the metastability of the Si-XIII phase and its main kinetic pathways linking it to the other Si phases.

\section{Conclusions}
\label{sec:conc}

In this work, we have resolved the long-standing question regarding the structural assignment of the Si-XIII phase, which was first observed over 20 years ago but has remained unidentified until now. Thanks to a convergent methodology, combining nanoindentation, comprehensive TEM/SAED analysis, and first-principles modeling, we have identified a triclinic $P\bar{1}$ structure with eight atoms per unit cell, crystallographically resemblant of the R8 metastable phase.

However, despite this geometric kinship, Si-XIII should not be regarded as merely a distorted variant of R8. Its equilibrium volume per atom is significantly larger, placing it much closer to the diamond phases, and its formation energy is intermediate between the R8/BC8 metastable phases and the stable diamond phases. Our PES exploration further confirms that it occupies a well-defined local minimum of the silicon energy landscape, with the R8$\,\to\,$Si-XIII transition representing the lowest-barrier pathways escaping the R8 energy basin. This kinetic picture is directly confirmed by the experiments. The dominant Si-XIII$\,\to\,$\emph{dc} transition, with a minimum barrier of only $90~\text{meV/atom}$, naturally explains the thermal instability of Si-XIII observed at $\sim$250$^{\circ}$C. Supplementary furnace annealing experiments corroborate this scenario, showing a clear reduction of Si-XIII Raman signatures and recovery toward the diamond phase, in agreement with the computed transition pathways. These  findings establish the proposed Si-XIII phase as a kinetically and thermodynamically distinct phase in its own right. 

The proposed structure achieves comprehensive agreement with all available experimental fingerprints of Si-XIII: interplanar spacings from multi-zone-axis SAED, characteristic Raman frequencies, and thermodynamic metastability that is fully consistent with the silicon phase diagram. To the best of our knowledge, this structure has not been previously reported in any theoretical or experimental study, representing a genuinely new silicon allotrope.

Beyond the specific case of Si-XIII, the challenge addressed here is a paradigmatic example of solid-solid phase transitions, demonstrating the power of integrating theoretical modeling with correlated experimental validation to resolve structural problems in complex materials systems.

\section*{CRediT authorship contribution statement}

\textbf{Fabrizio Rovaris:} Conceptualization, Methodology, Software,
Formal analysis, Investigation, Data curation, Visualization, Validation,
Writing -- original draft, Writing -- review \& editing.
\textbf{Corrado Bongiorno:} Conceptualization, Methodology, Formal analysis,
Investigation, Validation, Visualization, Writing -- original draft,
Writing -- review \& editing.
\textbf{Anna Marzegalli:} Conceptualization, Investigation, Validation,
Writing -- review \& editing.
\textbf{Mouad Bikerouin:} Investigation, Visualization.
\textbf{Davide Spirito:} Investigation, Visualization.
\textbf{Gerald J.K. Schaffar:} Investigation, Visualization.
\textbf{Mohamed Zaghloul:} Investigation.
\textbf{Agnieszka Anna Corley-Wiciak:} Investigation, Writing -- review \& editing.
\textbf{Francesco Montalenti:} Investigation, Writing -- review \& editing.
\textbf{Verena Maier-Kiener:} Conceptualization, Methodology, Formal analysis,
Investigation, Writing -- review \& editing.
\textbf{Giovanni Capellini:} Conceptualization, Investigation, Resources,
Supervision, Writing -- review \& editing.
\textbf{Antonio M. Mio:} Conceptualization, Methodology, Data curation,
Investigation, Resources, Validation, Visualization, Supervision,
Writing -- original draft, Writing -- review \& editing.
\textbf{Emilio Scalise:} Conceptualization, Methodology, Software,
Formal analysis, Investigation, Resources, Funding acquisition,
Project administration, Supervision, Validation, Visualization,
Writing -- original draft, Writing -- review \& editing.

\section*{Declaration of competing interests}

The authors declare that they have no known competing financial interests
or personal relationships that could have appeared to influence the work
reported in this paper.

\section*{Data availability}

The data that support the findings of this study are available from the
corresponding authors upon reasonable request. The crystal structure of
Si-XIII reported in this work (fractional coordinates and lattice parameters)
is provided in the Supplementary Material file of this manuscript.

\section*{Acknowledgements}

Fruitful discussions with Prof. Kenneth Beyerlein and Rasool Doostkam (Institut National de la Recherche Scientifique, Canada) are gratefully acknowledged. 

F.R., F.M. and E.S. acknowledge the CINECA consortium under the ISCRA initiative for the availability of high-performance computing resources and support.

F.R., M.B., E.S., A.M.M. and M.Z acknowledge financial support under the National Recovery and Resilience Plan (NRRP), Mission 4, Component 2, Investment 1.1, Call for tender No. 104 published on 2.2.2022 by the Italian Ministry of University and Research (MUR), funded by the European Union – NextGenerationEU – Project Title "SiGe Hexagonal Diamond Phase by nanoIndenTation (HD-PIT)", project number: 2022-NAZ-0098 – CUP H53D23000780001 and B53D23004120006 - Grant Assignment Decree No. 957 adopted on 30.06.2023 by the Italian Ministry of Ministry of University and Research (MUR). 

D.S. acknowledges financial support under the "Ramón y Cajal" programme funded by the Spanish MICIU/AEI/10.13039/501100011033 and ESF+ (grant no. RYC2022-037186-I).

Part of this work, has been carried out within the Joint Lab “Intelligent electrooptical sensing” established between IHP-Leibniz Institute for High Performance Microelectronics and Roma Tre University.

\bibliographystyle{unsrt}
\bibliography{biblio}

\end{document}


\maketitle

\begin{table}[htbp]
\centering
\setlength{\tabcolsep}{8pt}
\begin{tabular}{llccc}
\hline\hline
Atom & Wyckoff & $x$ & $y$ & $z$ \\
\hline
Si$_1$  & $2i$ & 0.014 & 0.492 & 0.689 \\
Si$_2$  & $2i$ & 0.986 & 0.508 & 0.311 \\
\hline
Si$_3$  & $2i$ & 0.170 & 0.943 & 0.974 \\
Si$_4$  & $2i$ & 0.830 & 0.057 & 0.026 \\
\hline
Si$_5$  & $2i$ & 0.653 & 0.685 & 0.163 \\
Si$_6$  & $2i$ & 0.347 & 0.315 & 0.837 \\
\hline
Si$_7$  & $2i$ & 0.380 & 0.826 & 0.318 \\
Si$_8$  & $2i$ & 0.620 & 0.174 & 0.682 \\
\hline\hline
\end{tabular}
\caption{Crystal structure of Si-XIII. Space group $P\bar{1}$ (No.~2),
         triclinic system, point group $\bar{1}$.
         All 8 atomic positions (fractional coordinates) of the
         primitive unit cell are listed; atoms are grouped in pairs
         related by the inversion center of the $2i$ Wyckoff site.}
\label{tab:SiXIII_structure}
\end{table}

\begin{table}[t]
\setlength\tabcolsep{2.5pt}
\small
  \makebox[\linewidth][c]{%
 \begin{tabular}{lcccccccccl}
    \hline
    Phase & Approx. & $a$ [\AA] & $b$ [\AA] & $c$ [\AA] & $\alpha$ [$^{\circ}$] & $\beta$ [$^{\circ}$] & $\gamma$ [$^{\circ}$] & $V$  & $\Delta E$\\
    (Space group)  &   &   &   &   &   &   &   &  [\AA$^{3}/\text{atom}$]&[meV/atom]\\
    \hline
    \multirow{5}{*}{\emph{dc} (Fd$\overline{3}$m)}  & LDA    & 3.805 & 3.805 & 3.805 & 60.0 & 60.0 & 60.0 & 19.48  & 0 \\
                                & PBEsol & 3.842 & 3.842 & 3.842 & 60.0 & 60.0 & 60.0 & 20.05  & 0 \\
                                & PBE    & 3.861 & 3.861 & 3.861 & 60.0 & 60.0 & 60.0 & 20.35  & 0 \\
                                & SCAN   & 3.839 & 3.839 & 3.839 & 60.0 & 60.0 & 60.0 & 20.01  & 0 \\
                                & GAP    & 3.861 & 3.861 & 3.861 & 60.0 & 60.0 & 60.0 & 20.36  & 0 \\
                                & Lit.~\cite{Mohr2014}   & 3.840 & 3.840 & 3.840 & 60.0 & 60.0 & 60.0 & 20.03  & -- \\
    \hline
    \multirow{5}{*}{R8 (R$\overline{3}$)}         & LDA    & 5.671 & 5.671 & 5.671 & 109.9 & 109.9 & 109.9 & 17.29  & 112 \\
                                & PBEsol & 5.720 & 5.720 & 5.720 & 109.9 & 109.9 & 109.9 & 17.73  & 116 \\
                                & PBE    & 5.759 & 5.759 & 5.759 & 109.8 & 109.8 & 109.8 & 18.12  & 158 \\
                                & SCAN   & 5.749 & 5.749 & 5.749 & 109.8 & 109.8 & 109.8 & 18.04  & 205 \\
                                & GAP    & 5.749 & 5.749 & 5.745 & 109.6 & 109.6 & 109.6 & 18.16  & 175 \\
                                & Exp.   & 5.620  & 5.620  & 5.620  & 109.8 & 109.8 & 109.8   &  16.87 & -- \\ 
                                & Lit~\cite{PiltzPRB1995} (at 8.2 Gpa) & 5.620 & 5.620 & 5.620 & 110.1 & 110.1 & 110.1  &  16.69 & -- \\
    \hline
    \multirow{5}{*}{BC8 (Ia$\overline{3}$)}& LDA    & 5.670 & 5.670 & 5.670 & 109.5 & 109.5 & 109.5 & 17.54 & 124 \\
                                & PBEsol & 5.720 & 5.720 & 5.720 & 109.5 & 109.5 & 109.5 & 18.00  & 119 \\
                                & PBE    & 5.758 & 5.758 & 5.758 & 109.5 & 109.5 & 109.5 & 18.37  & 157 \\
                                & SCAN   & 5.747 & 5.747 & 5.747 & 109.5 & 109.5 & 109.5 & 18.26  & 186 \\
                                & GAP    & 5.754 & 5.754 & 5.754 & 109.5 & 109.5 & 109.5 & 18.33  & 165 \\
                                & Lit.~\cite{PiltzPRB1995} & 5.748 & 5.748 & 5.748 & 109.5 & 109.5 & 109.5  &  18.28 & -- \\
    \hline
    \multirow{5}{*}{Si-XIII (P$\bar{1}$)}    & LDA    & 5.150 & 5.745 & 6.204 & 109.1 & 100.0 & 107.1 & 19.78  & 98 \\
                                & PBEsol & 5.200 & 5.797 & 6.263 & 109.1 & 100.0 & 107.0 & 20.35  & 94 \\
                                & PBE    & 5.225 & 5.828 & 6.297 & 109.0 & 100.0 & 107.0 & 20.70  & 98 \\
                                & SCAN   & 5.191 & 5.807 & 6.266 & 109.1 & 100.0 & 107.0 & 20.38  & 116 \\
                                & GAP    & 5.186 & 5.796 & 6.385 & 109.0 & 100.8 & 106.3 & 20.72  & 128 \\
                                & Exp.    & 5.20  & 5.74  & 6.21  & 109   & 100   & 107   &  20.00 & - \\ 
    \hline    
  \end{tabular}%
  }
\caption{Equilibrium lattice parameters of the proposed Si-XIII structure obtained from DFT relaxations with different exchange-correlation functionals and the GAP interatomic potential, compared to the metastable R8 and BC8 phases and the reference \emph{dc}-Si. The atomic volume and cohesive energy are also reported, showing that the Si-XIII phase lies closer to the stable \emph{dc}-Si, both in terms of atomic volume and, more notably, in cohesive energy. For the \emph{dc}, R8, and BC8 phases values reported in the literature are reported together with the references they were taken from. For the Si-XIII and R8 phases, cell parameters estimated on the basis of the experimental SAED patterns are also reported. Note that the current estimated values for the R8 phase suggest a possible deformation of the R8 structure due to residual strain, in agreement with previous observations of the R8 phase obtained by nanoindentation~\cite{wong2019formation}.}

 \label{tab:cell_params_supp}
\end{table}

\begin{table}
    \makebox[\linewidth][c]{%
    \begin{tabular}{cccc|cccc}
    \hline
    \multicolumn{4}{c}{Si-XIII (P$\bar{1}$)} & \multicolumn{4}{c}{R8} \\
    \emph{hkl} & simulated & measured & deviation & \emph{hkl} & simulated & measured & deviation\\
    \hline
    (001) & 5.606 & 5.62 & 0.014 (0.2\%) & (-110) & 4.598 & 4.54 & -0.058 (-1.3\%) \\ 
    (-110) & 4.365 & 4.35 & -0.015 (-0.3\%) & (0-11) & 4.598 & 4.59 & -0.008 (-0.2\%) \\ 
    (1-11) & 3.573 & 3.54 & -0.033 (-0.9\%) & (-101) & 4.598 & 4.59 & -0.008 (-0.2\%) \\ 
    (-111) & 3.328 & 3.33 & 0.002 (0.1\%) & (-12-1) & 2.655 & 2.64 & -0.015 (-0.6\%) \\ 
    (101) & 3.186 & 3.18 & -0.006 (-0.2\%) & (-211) & 2.655 & 2.64 & -0.015 (-0.6\%) \\ 
    (011) & 3.168 & 3.16 & -0.008 (-0.3\%) & (-1-12) & 2.655 & 2.66 & 0.005 (0.2\%) \\ 
    (1-12) & 2.441 & 2.42 & -0.021 (-0.9\%) & (0-12) & 2.655 & 2.66 & 0.005 (0.2\%) \\ 
    (-112) & 2.284 & 2.3 & 0.016 (0.7\%) & (-102) & 2.655 & 2.66 & 0.005 (0.2\%) \\ 
    (2-11) & 2.223 & 2.22 & -0.003 (-0.1\%) & (-221) & 2.051 & 2.03 & -0.021 (-1\%) \\ 
    (-121) & 2.205 & 2.2 & -0.005 (-0.2\%) & (1-22) & 2.051 & 2.03 & -0.021 (-1\%) \\ 
    (102) & 2.15 & 2.16 & 0.01 (0.5\%) & (-1-13) & 1.867 & 1.88 & 0.013 (0.7\%) \\ 
    (012) & 2.116 & 2.14 & 0.024 (1.1\%) & (0-23) & 1.735 & 1.75 & 0.015 (0.9\%) \\ 
    (2-21) & 2.086 & 2.06 & -0.026 (-1.2\%) & (-203) & 1.735 & 1.75 & 0.015 (0.9\%) \\ 
    (021) & 2.014 & 2.03 & 0.016 (0.8\%) & (1-33) & 1.452 & 1.45 & -0.002 (-0.1\%) \\ 
    (-221) & 1.985 & 1.97 & -0.015 (-0.8\%) & (-313) & 1.452 & 1.45 & -0.002 (-0.1\%) \\ 
    (201) & 1.984 & 1.98 & -0.004 (-0.2\%) &    &    &  &  \\ 
    (2-12) & 1.784 & 1.75 & -0.034 (-1.9\%) &    &    &  &  \\ 
    (1-13) & 1.765 & 1.72 & -0.045 (-2.5\%) &    &    &  &  \\ 
    (-122) & 1.726 & 1.75 & 0.024 (1.4\%) &    &    &  &  \\ 
    (-131) & 1.581 & 1.59 & 0.009 (0.6\%) &    &    &  &  \\ 
    (3-11) & 1.552 & 1.54 & -0.012 (-0.8\%) &    &    &  &  \\ 

    \hline
    \end{tabular}%
    }
    \caption{Measured and simulated \textit{d}-spacings for each (\emph{hkl}) observed reflection of Si-XIII and R8. Simulated values are obtained using the experimental lattice parameters reported in Table \ref{tab:cell_params_supp}. For each reflection, the deviation between measured and simulated d-spacings is reported both as a value and as a percentage.}
    \label{tab:placeholder}
\end{table}

\clearpage

\begin{figure*}[htb]
    \centering
    \includegraphics[width=1.0\linewidth]{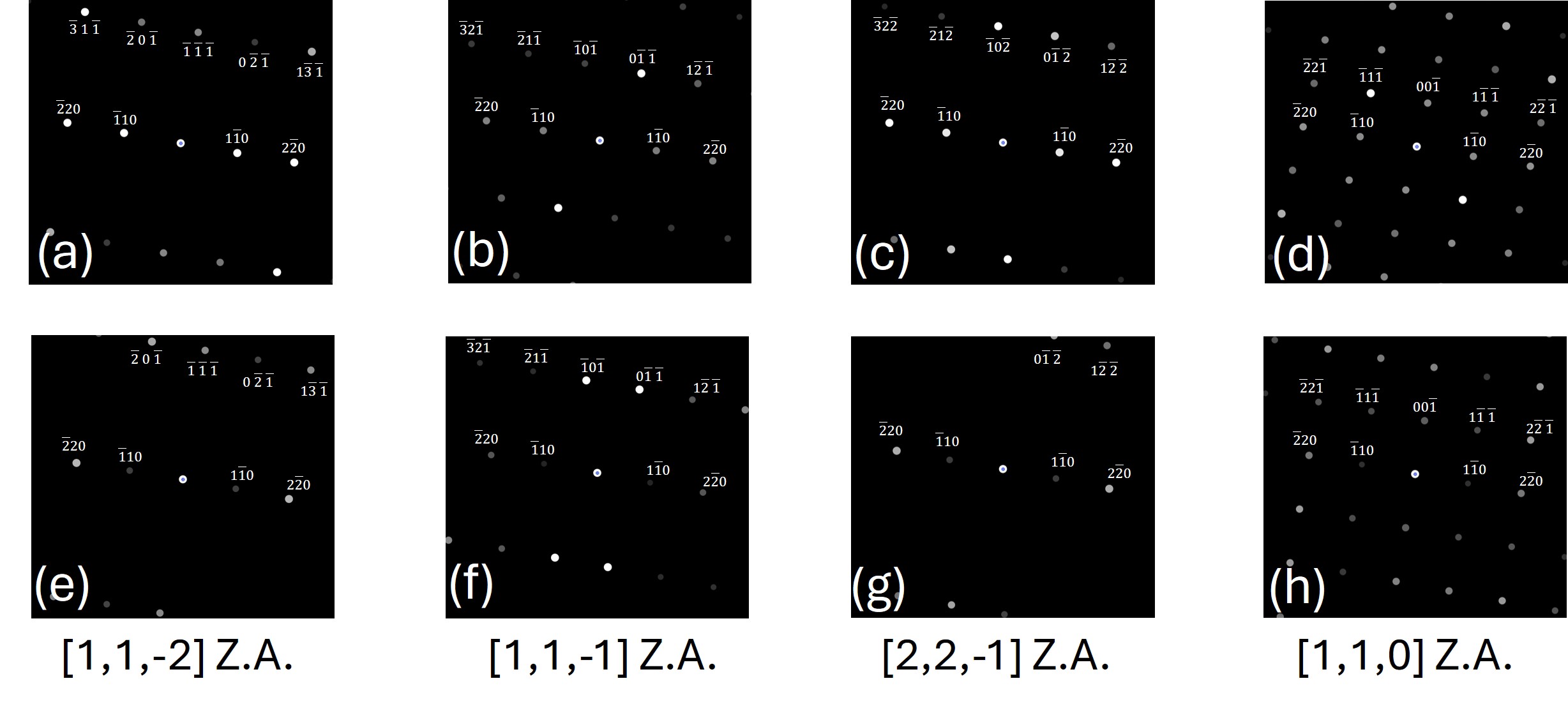}
    \caption{(a-d) Simulated SAED of the Si-XIII phase across multiple zone axes, as reported in Fig. 1(h-k). For comparison, the corresponding simulated SAED of the R8 phase is shown in panels (e-h). Indexing of the R8 phase is presented in the rhombohedral basis, as defined in Table~\ref{tab:cell_params_supp} }
    \label{fig:saedSI}
\end{figure*}

\bibliographystyle{unsrt}
\bibliography{biblio}